\documentclass[titlepage,oneside,12pt,letter,fleqn,numbered]{article}

\usepackage[tiny,rm]{titlesec}
\newpagestyle{trbstyle}{
	\sethead{}{}{\thepage}
}
\titleformat{\section}{\bfseries}{}{0pt}{\uppercase}
\titlespacing*{\section}{0pt}{12pt}{*0}
\titleformat{\subsection}{\bfseries}{}{0pt}{}
\titlespacing*{\subsection}{0pt}{12pt}{*0}
\titleformat{\subsubsection}{\itshape}{}{0pt}{}
\titlespacing*{\subsubsection}{0pt}{12pt}{*0}

\usepackage{enumitem}
\setlist[1]{labelindent=0.5in,leftmargin=*}
\setlist[2]{labelindent=0in,leftmargin=*}

\usepackage{ccaption}
\usepackage[doublespacing]{setspace}
\usepackage{amsmath}
\usepackage{amsfonts}
\usepackage{booktabs}
\usepackage[table,xcdraw]{xcolor}
\makeatletter
\renewcommand{\fnum@figure}{\textbf{FIGURE~\thefigure} }
\renewcommand{\fnum@table}{\textbf{TABLE~\thetable} }
\makeatother
\captiontitlefont{\bfseries \boldmath}
\captiondelim{\;}

\usepackage[sort,numbers,round]{natbib}

\setcitestyle{round}

\usepackage[pagewise,mathlines,displaymath]{lineno}

\usepackage{totcount}
	\regtotcounter{table} 	
	\regtotcounter{figure} 	

\newcommand\wordcount{
    \immediate\write18{texcount -sum -1 \jobname.tex > 'count.txt'} \input{count.txt} }

\usepackage{mathptmx}
\usepackage{times}
\usepackage{cases}
\usepackage{textcomp}
\usepackage{subfig}
\usepackage{graphicx}
\graphicspath{{figures/}}
\usepackage{url}
\usepackage{epstopdf}
\usepackage{color}
\usepackage{mathrsfs,MnSymbol}
\usepackage{booktabs}
\usepackage{multirow}
\usepackage{soul,color}
\usepackage{enumitem}
\usepackage{url}
\usepackage{cleveref}

\usepackage{caption}
\usepackage{float}
\usepackage{placeins}

\begin{document}

\thispagestyle{empty}

\begin{titlepage}
\begin{flushleft}

{\uppercase{ \textbf{Examining the interactions between working from home, travel behavior and change in car ownership due to the impact of COVID-19}}
}\\[1cm]

  \textbf{Yunhan Zheng (corresponding author)}\\
  Department of Civil and Environmental Engineering\\ Massachusetts Institute of Technology, Cambridge, MA 20139\\
  Email: yunhan@mit.edu; ORCiD: 0000-0001-5114-7561\\[0.5cm]

  \textbf{Nicholas Caros}\\
  Department of Civil and Environmental Engineering\\
  Massachusetts Institute of Technology, Cambridge, MA 20139\\
  Email: caros@mit.edu\\[0.5cm]

\textbf{Jim Aloisi}\\
  Department of Urban Studies and Planning\\
  Massachusetts Institute of Technology, Cambridge, MA 02139 \\
  Email: jaloisi@mit.edu\\[0.5cm]

\textbf{Jinhua Zhao}\\
  Department of Urban Studies and Planning\\
  Massachusetts Institute of Technology, Cambridge, MA 02139 \\
  Email: jinhua@mit.edu; ORCiD: 0000-0002-1929-7583\\[0.5cm]

		    \textbf{Declarations of interest:} none. \\
    \textbf{Acknowledgement:} this work was supported by the MIT Energy Initiative’s Mobility of the Future Study and the Barr Foundation.



\end{flushleft}
\end{titlepage}

\newpage

\thispagestyle{empty}

\begin{center}
\large{\textbf{Examining the interactions between working from home, travel behavior and change in car ownership due to the impact of COVID-19}}\\[1cm]
\end{center}
\section*{Abstract}

COVID-19 has disrupted society and changed how people learn, work and live. The availability of vaccines in the spring of 2021, however, led to a gradual return of many pre-pandemic activities in Massachusetts in the fall of 2021. Leveraging data that were collected using a map-based survey tool in the Greater Boston area in the fall of 2021, this study explores changes in travel behavior due to COVID-19 and investigates the underlying factors contributing to these changes. First, a structural equation modeling technique is developed to capture the interactions between various travel choices, including working from home, travel mode use and change in car ownership. Moreover, attitudinal factors such as risk perceptions and attitudes towards WFH are incorporated into the framework to explain behavior changes. Second, a discrete choice modeling approach is taken to study shifts in commuting mode choices in the fall of 2021. The results show that in the fall of 2021, people became more likely to use their cars to commute, and for those who bought cars during the pandemic, they tended to work on-site more. Our findings can provide planners and policymakers with information upon which to base travel demand management decisions in the post-pandemic era.

\bigskip
\noindent
\emph{Keywords}: COVID-19, Remote work, Commuter mode choice, Travel behavior, Car ownership
\newpage

\section{1. Introduction}





The COVID-19 pandemic has dramatically altered personal mobility choices and social activities. During the pandemic, there has been a major shift towards work from home (WFH) arrangements \cite{mouratidis2021covid,shakibaei2021impact}, and travel behavior have also undergone profound changes. Effective vaccines, new anti-viral medications and general population fatigue associated with the pandemic have combined to gradually shift larger numbers of people toward normal daily routines. It is now particularly useful  to understand to what degree the behavioral shifts that occurred during the COVID-19 period will continue as COVID-19 restrictions are lifted and society gradually moves back towards normal activity patterns. Understanding people's behavior changes and the underlying factors behind these changes in the post-pandemic period could help public transportation operators and policymakers identify factors that could be targeted if certain travel choices are to be encouraged. In addition to traditional travel behavior predictors such as socio-demographic and built environment features, new factors such as risk-taking attitudes and perceptions towards WFH should also be considered when predicting people's travel behavior shifts. As such, this research develops a comprehensive model that captures the simultaneous effects of risk-taking perceptions, attitudes towards WFH, the number of in-person work days, car ownership changes and travel behavior. Given that WFH has made a profound impact on travel demand, we develop a separate model to study factors that govern people's WFH choices and their satisfactions towards WFH. \\

\noindent We collected survey data about  travel choices and attitudes of residents in Greater Boston area during the fall of 2021, a time when COVID-19 restrictions were largely lifted and people were gradually returning to the office. By estimating models of respondents' travel behavior during both the pre-pandemic and the fall of 2021, this study aims to understand the impacts of COVID-19 on travel behavior and attitudes in the fall of 2021. Though a great number of previous studies have investigated the impacts of the pandemic on travel demand during the pandemic \cite{benita2021human,abdullah2020exploring,mo2021modeling}, research into travel demand in a period when the COVID-19 restrictions are lifted is scarce. For those that focused on the post-pandemic travel shifts, most studies collected data on self-reported expectations of travel behavior in the post-pandemic period. This study is among the very few studies that examine the factors that impact travel choices during the fall of 2021 in a U.S. city using retrospective travel behavior and attitudinal data. \\

\noindent A three-pronged approach is used herein to elicit the factors that influence travel behavior and WFH during the fall of 2021. First, we use a structural equation modeling approach to study: 1) the interactions among different travel choices including usage frequencies for private cars, transportation network companies (TNCs) like Uber and Lyft, public transit (PT), and walking, as well as changes in car ownership and the frequency of on-site working; 2) how perceived risks of contracting COVID-19 when using different travel modes affect travel behavior. We identify how the relationships among these key variables have changed due to COVID-19 by comparing the results both before the pandemic and during the fall of 2021. While there has been literature exploring the relationship between WFH and travel behavior during the COVID-19 pandemic, we propose a holistic conceptual framework that highlights the interrelationships among travel behavior, WFH, car ownership and travel related attitudes, which is then tested using a structural equation model (SEM). \\

\noindent Second, we construct a separate SEM model to study factors that contribute to WFH choices and overall satisfaction with WFH. Attitudinal factors including satisfaction with WFH along several dimensions such as work productivity and work-life balance, as well as risk perceptions of different travel modes, are considered. The knowledge of these attitudinal factors is important to design policy measures to encourage positive behavior and social outcomes \cite{jain2022covid}.\\

\noindent Third, a discrete choice modeling approach is estimated to predict respondents' primary commuting modes. As WFH is likely to persist to some degree post-COVID, we incorporate WFH into the mode choice set along with the traditional travel mode alternatives (auto, active and PT modes). In addition to commonly collected survey data such as the socio-demographic data, a map-based survey tool was incorporated into the survey to gather locations of respondents' residential locations and commuting trips. This survey design allows for the collection of mode-specific attributes (\textit{e.g.} travel time) and built environment variables (\textit{e.g.} distance to the central business district, or CBD). This study draws on these rich data sources to assess the impact of COVID-19 on respondents' commuting choices by estimating discrete choice models for both the pre-pandemic period and the fall of 2021. \\

\noindent The remaining sections of the paper are organized as follows. Section \ref{sec:literature} reviews previous literature related to the impacts of COVID-19 on travel behavior. Section \ref{sec:background} provides background information and a timeline of the major events surrounding the COVID-19 pandemic in Massachusetts to contextualize the survey results. Section \ref{sec:method} describes the data and methods used in this study. The empirical results and estimated models are presented in Section \ref{sec:results}, followed by a discussion and conclusions in Section \ref{sec:conclusion}.

\section{2. Literature Review}
\label{sec:literature}



This study builds on decades of research into travel behavior that preceded the pandemic, as well as the many papers that have explored the effect of the COVID-19 pandemic on travel across the globe. 
Existing literature in both fields are reviewed in this section.
The recent studies that investigate changes in travel behavior as a result of the pandemic are then summarized and the research gaps addressed by this paper are identified. \\

\noindent The attitudinal factors of travel behavior have been a rich area of research for many years \cite{levin1977measurement, zheng2022gender,zheng2021user}.
Structural equation modeling, used in this paper, is a popular method for eliciting relationships between observed travel behavior and a large number of endogenous, exogenous and latent variables \cite{golob2003structural}.
Dozens of studies have used structural equation modeling to determine the impact of different factors on travel behavior, including land use characteristics \cite{van2007effects}, home location \cite{bagley2002impact,de2021changes} and lifestyle \cite{etminani2015modeling}. 
These methods have been used to model car ownership, similar to the approach taken in this paper \cite{moody2020travel}. \\ 

\noindent The COVID-19 pandemic has had a tremendous effect on travel demand and patterns. There have been numerous studies quantifying the magnitude and direction of the changes in travel demand. 
A number of studies used aggregated data from mobile phones to compare pandemic travel patterns to pre-pandemic travel patterns in the United States and across countries \cite{gao2020mapping, heiler2020country, xiong2020mobile}.
\citet{benita2021human} provides an excellent review of the literature on mobility patterns during the pandemic. 
Surveys issued since the onset of the pandemic have found a measurable shift towards remote work as a substitute for commuting in Chicago \cite{shamshiripour2020covid} and the United States more broadly \cite{barrero2021working}. 
Other surveys about future intentions find that many commuters expect to shift away from public transit towards personal cars after the pandemic has subsided \cite{currie2021evidence}. \\

\noindent More recently, researchers have become interested in how the COVID-19 pandemic has affected decision-making when it comes to travel. 
Many of these discrete decisions such as mode choice are modeled using logistic regression (logit). 
Mode shifts away from shared modes (\textit{e.g.} public transit, taxis) has been observed across a wide range of studies due to concern about COVID-19 \cite{abdullah2021measuring, torrisi2021continue}. 
\citet{anwari2021exploring} finds that in Bangladesh, the lack of internet penetration in rural areas results in a limited reduction in work trips during the pandemic. \citet{parady2020travel} uses several different statistical models to test the direct effects of attitudes on trip frequencies, discovering that a low opinion of the self-control of others is associated with a reduction in activity frequency for several activity categories. 
A logistic regression model developed by \citet{abdullah2020exploring} explores the socioeconomic and attitudinal factors that affect mode choice before and during the pandemic, but does not include car ownership or frequency of remote work. 
Finally, \citet{beck2020slowly} estimates an ordered logit model for the choice of remote work days using survey data from Spring 2020, a relatively early period in the pandemic.\\

\noindent Integrated choice and latent variable (ICLV) discrete choice models have also been used to determine whether unobserved latent factors contribute to travel and work decisions during the COVID-19 pandemic. 
\citet{barbour2021statistical} issued a survey during the summer of 2020 to estimate two binary models, one for the choice of switching from in-person to remote work, and the other for the choice of continuing in-person work after COVID-19. 
Taking a similar approach, \citet{aaditya2021psychological} explores the role of latent influences in post-pandemic mode choice in India. 
Interestingly, the study finds that awareness of COVID-19 and perception of pandemic lockdown policies have a significant impact on mode choice. 
\citet{chen2021role} produced similar results using an ICLV model estimated from a 2020 survey of Dutch residents.\\

\noindent Several studies of pandemic travel behavior also use structural equation models.
Using a survey of Indonesian residents, \citet{irawan2021exploring} explores the effects of attitudes, norms and remote work on changes in the frequency travel and ride-hailing use during the pandemic. 
The authors found that a relaxed attitude towards the risk of COVID-19 was predictive of a smaller reduction in daily travel during the pandemic. 
\citet{javid2021structural} and \citet{elias2021extent} use structural equation models to uncover relationships between attitudes towards the spread of COVID-19 on public transit and the frequency of public transit use during the pandemic in Pakistan and Israel, respectively.
By examining longitudinal data, \citet{jiao2021measuring} finds that a rise in COVID-19 cases in Houston, Texas is predictive of less aggregate travel during the following week. 
Another interesting study uses structural equation modeling to identify the attitudinal and social determinants of preferences for post-COVID remote work among Australians \cite{jain2022covid}.\\

\noindent There have also been many studies of remote work and commuting focused on the United States context since the onset of the COVID-19 pandemic.
\citet{asmussen2022modeling} uses a stated-preference model to quantify the factors that impact workplace location choice by remote workers in Texas, finding that workplace environment is at least as important as the geographic location of the workplace.
\citet{salon2021potential} reviewed whether observed changes in a wide range of travel-related choices, including air travel, home relocation and commuting among U.S. remote workers after the pandemic were likely to persist over time. 
Taking a slightly different approach, \citet{tahlyan2022longitudinal} uses data from a longitudinal survey of remote workers across the United States to determine the direction and magnitude of changes in commuting patterns and other travel behavior during different stages of the pandemic.
Lastly, survey data in the Seattle area has been used to identify clusters of remote workers based on similar attitudes towards remote work and intention to continue remote work in the future \cite{kong2022work}. 
These studies provide helpful insights into travel behavior in the United States, but there remain no comprehensive studies of the relationship between remote work, mode choice and car ownership across multiple time periods. \\

\noindent This paper builds on previous research in three important ways. 
First, the study focuses on the Greater Boston area, an important global hub for education, healthcare and biotechnology.
Second, it develops a comprehensive model of three different travel factors (work from home frequency, mode choice and car ownership) rather than looking at each factor separately as previous studies in the United States and elsewhere have done.
Finally, this paper estimates these comprehensive models of travel behavior both before and after the pandemic, allowing comparison between the contributing factors during baseline period and the fall of 2021. 

\section{3. Background}
\label{sec:background}
This paper is focused on understanding the factors that contributed to travel behavior of residents of the Commonwealth of Massachusetts in the United States.
To provide context for the analysis and results, this section reviews the timeline of major events surrounding the outbreak of the COVID-19 in Massachusetts and the response by the public and private sectors. 
The first case of COVID-19 in Massachusetts was confirmed in early February 2020 \cite{massdph2020}.
A corporate gathering at the end of February 2020 in Boston, the largest city in Massachusetts, was one of the first known events to result in significant spread of the virus in the United States \cite{arsenault2020biogen}. 
In early March 2020, the virus was beginning to spread quickly, leading to sweeping school and office closures \cite{reiss2020schools}. 
So-called ``non-essential businesses'' were ordered to close on March 23, 2020 and a stay-at-home order was issued by the Governor \cite{markos2021lockdown}. \\

\noindent A reopening plan was gradually implemented throughout the summer of 2020 \cite{hilliard2020reopening}, but an increase in COVID-19 cases in the fall of 2020 resulted in a tightening of some restrictions on businesses and gatherings \cite{stout2020mandate}.
The availability of vaccines in the spring of 2021 and relatively high vaccination rates in Massachusetts led to a gradual return of many pre-pandemic activities in the summer and fall of 2021.
Restrictions on dining, masks and indoor gatherings were lifted \cite{platoff2021policies}.
All major universities and public schools resumed full time in-person teaching in September 2021 \cite{carlin2021college}. 
Commercial districts were seeing employers return to in-office work  \cite{gardizy2021office}, and public transit ridership in Boston continued to recover \cite{bleichfield2021mbta}. This paper focuses on the fall of 2021 as the study period.

\section{4. Data and Methods}
\label{sec:method}
This study uses an online survey to collect data about perceptions towards work from home, their travel behavior and preferences, as well as their car ownership. The participant recruitment, distribution platform and survey design are described in the subsections that follow.

\subsection{4.1 Participants}
The survey was designed as a 15-minutes online survey with the Qualtrics platform. The survey was distributed from October 25 to November 8, 2021 and December 8, 2021 to December 18, 2021 through Qualtrics, Prolific and MTurk recruitment panels. Data collection was paused from November 8, 2021 to December 8, 2021 to avoid the Thanksgiving holiday, when many Americans are on holiday. Our study employed a mixed sampling strategy, including quota sampling and convenience sampling. Specifically, we set quotas for gender and race to ensure that the sample distribution reflects the population distribution on these dimensions, while also allowing participants to voluntarily choose to participate in the survey. The use of quota sampling allowed us to achieve a sample that was more reflective of the population, while convenience sampling provided a practical and cost-effective means of recruitment. \\ 


\noindent The survey was distributed in the fall of 2021, during which restrictions on dining, masks and indoor gatherings were lifted and commercial districts were seeing employers return to in-office work as mentioned above. Therefore, analyzing travel behavior and preferences during this period can provide insights into the long-term travel impacts of COVID-19. \\

\begin{figure}[!t]
    \centering
    \includegraphics[width=6 in]{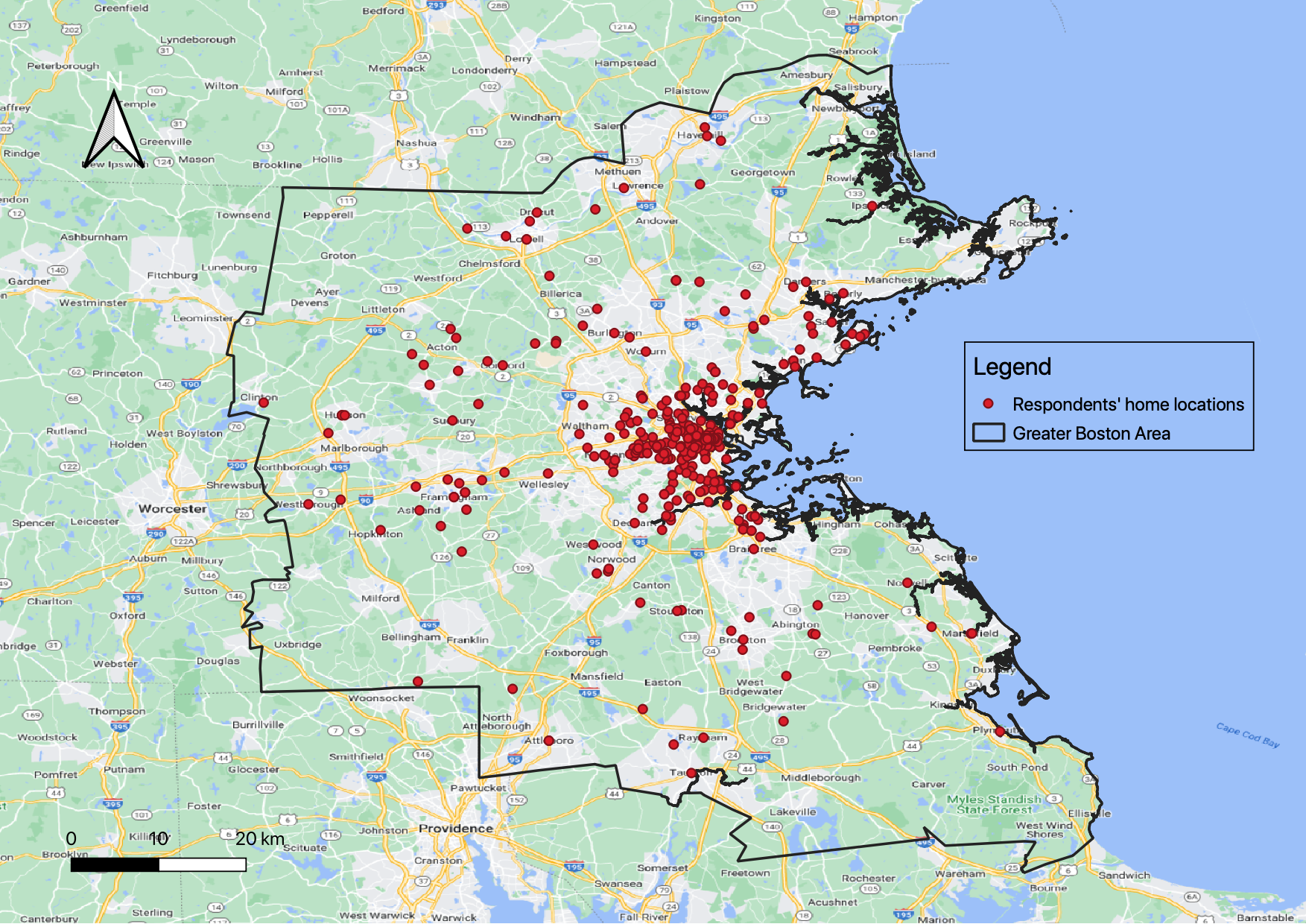}
    \caption{Map of the Greater Boston area showing the approximate residential locations of survey participants in the fall of 2021}
    \label{fig:respondent_loc}
\end{figure}

\noindent In the end, a total of 498 responses were collected from across Massachusetts. We selected participants whose home locations in the fall of 2021 are in the Greater Boston area, which leaves us a total of 425 samples. The descriptive statistics of the sample are summarized in Tables \ref{tab_desc} and \ref{tab_desc2}. Figure \ref{fig:respondent_loc} shows the approximate residential locations of the survey participants in the fall of 2021.

\subsection{4.2 Questionnaire}
The survey includes questions related to respondents' work status and travel patterns (for the pre-pandemic and the fall of 2021), their attitude towards WFH, and their socio-demographic information as presented in Tables \ref{tab_desc} and \ref{tab_desc2}.\\

\noindent This study aims to model travel patterns in both the pre-pandemic period and the fall of 2021 along two related dimensions: the frequency of use for different travel modes and commuting mode choices. For the former, we measure the travel mode usage frequencies with the question ``[Currently/Before COVID], on average, how frequently do you use each of the following travel modes?'' 
For the commuting mode choice prediction, we study the choice of primary travel mode for commuting to work/school among four categories of alternatives: 1) auto (including car driver, car passenger and taxi or ride hailing); 2) public transport (including bus, subway and commuter rail); 3) active modes (including walking and biking); 4) WFH.\\

\begin{figure}
\centering
\subfloat[Pre-pandemic period]{\label{fig:pre_map}\includegraphics[width=3in]{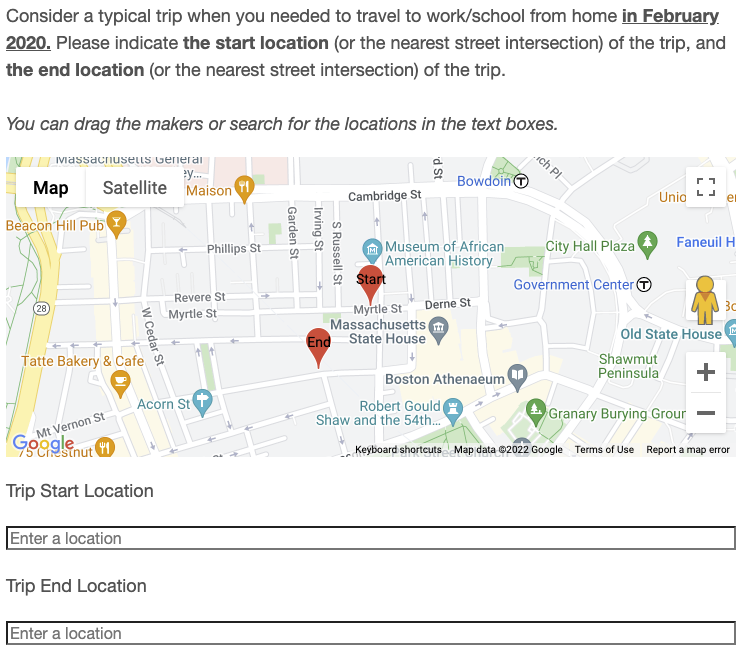}} 
\qquad
\subfloat[Fall 2021]{\label{fig:now_map}\includegraphics[width=3in]{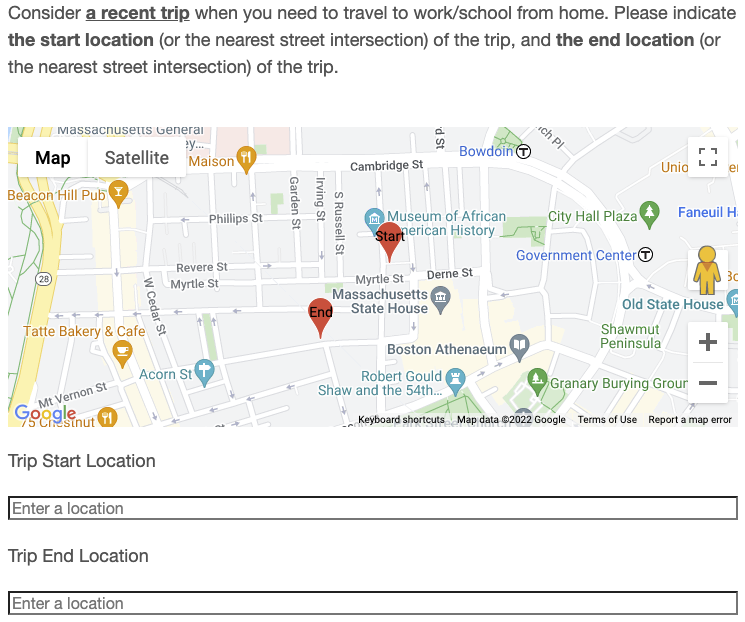}}
\caption{The online interface of the Google Maps survey tool used to collect the origin and destination of respondents' typical commuting trips} 
\label{fig_map} 
\end{figure}

\noindent To capture the trip-specific attributes that help predict commuting mode choices, we integrated a Google Maps interactive module into the survey. We asked the respondents to indicate the start and end locations of a typical commuting trip on the map, based on which a Google Maps API call yields the distance and travel time by different transport modes. Figure \ref{fig_map} shows the online interface which asks the respondents to mark the origin and destination of their typical commuting trips in the pre-pandemic and the fall of 2021. For those who chose WFH, we asked them to indicate the origin and destination of a typical commuting trip once the COVID-19 ceases to be a pandemic, which enabled us to calculate their travel time for other alternatives (\textit{i.e.} active, auto and transit). To ensure the accuracy of our data on trip start and end locations, we employed several strategies. First, we excluded samples where respondents reported the same start and end locations for a trip, as these individuals may not have been paying close attention to the survey questions. Additionally, we cross-referenced the reported trip start locations with respondents' home zip codes to confirm the accuracy of the data. Furthermore, since our study focuses specifically on travel behaviors within the Greater Boston area, we excluded individuals who reported traveling outside of this geographic region. By limiting our analysis to a specific area, we can more accurately capture the unique travel patterns and behaviors of this population.\\




\noindent To understand how attitudes towards working from home affect WFH behavior and travel behavior, we asked the respondents to evaluate WFH in comparison with in-person work using 5-point Likert style statements. In addition to asking the respondents to indicate their general satisfaction at work during WFH compared with in-person work, we also measured their attitudes towards WFH along several dimensions such as work productivity, creativity at work, work environment and work-life balance. Job requirements are also expected to affect WFH behavior, therefore three related questions were included in the survey: total number of working days, whether the job must be conducted at a work site, and whether the employer reimburses the individual for any portion of WFH costs.\\

\noindent One important factor that might affect usage of different travel modes is the perceived risk of becoming infected with COVID-19 at the time the survey was conducted. In light of this, we used a 5-point Likert scale question to measure perceived risk of catching the virus for each travel mode, and incorporated this question into the model estimations for the usage frequencies of different travel modes and perceptions towards WFH. 

\FloatBarrier
\begin{table}[h]
  \caption{Descriptive statistics}
  \label{tab_desc} 
\resizebox{\textwidth}{!}{%
\begin{tabular}{llllll}
\toprule
Variable                                     & Description                                                                                                                                                                                                                                                                            & Mean                  & Std.dev               & Min                 & Max                \\ \midrule
\multicolumn{6}{l}{\textit{\textbf{1. Travel behavioral   variables}}}                                                                                                                                                                                                                                                                                                                                                           \\\\[-1.8ex]
Commute mode: active   (pre-pandemic)        & \multirow{8}{*}{Primary   commuting mode (0/1)}                                                                                                                                                                                                                                        & 0.10                  & 0.31                  & 0                   & 1                  \\
Commute mode: active (fall   2021)           &                                                                                                                                                                                                                                                                                        & 0.10                  & 0.30                  & 0                   & 1                  \\
Commute mode: transit   (pre-pandemic)       &                                                                                                                                                                                                                                                                                        & 0.22                  & 0.42                  & 0                   & 1                  \\
Commute mode: transit (fall   2021)          &                                                                                                                                                                                                                                                                                        & 0.20                  & 0.40                  & 0                   & 1                  \\
Commute mode: auto   (pre-pandemic)          &                                                                                                                                                                                                                                                                                        & 0.63                  & 0.48                  & 0                   & 1                  \\
Commute mode: auto (fall 2021)               &                                                                                                                                                                                                                                                                                        & 0.56                  & 0.50                  & 0                   & 1                  \\
Commute mode: WFH   (pre-pandemic)           &                                                                                                                                                                                                                                                                                        & 0.04                  & 0.19                  & 0                   & 1                  \\
Commute mode: WFH (fall 2021)                &                                                                                                                                                                                                                                                                                        & 0.14                  & 0.35                  & 0                   & 1                  \\\hline\\[-1.8ex]
Travel time   (mins): active (pre-pandemic)  & \multirow{6}{*}{Travel time by different commuting modes in minutes}                                                                                                                                                                                                                   & 18.39                 & 18.74                 & 3                   & 114                \\
Travel time (mins): active   (fall 2021)     &                                                                                                                                                                                                                                                                                        & 21.39                 & 23.17                 & 3                   & 117                \\
Travel time (mins): transit   (pre-pandemic) &                                                                                                                                                                                                                                                                                        & 38.10                 & 19.35                 & 5                   & 91                 \\
Travel time (mins): transit   (fall 2021)    &                                                                                                                                                                                                                                                                                        & 37.44                 & 18.75                 & 9                   & 89                 \\
Travel time (mins): auto   (pre-pandemic)    &                                                                                                                                                                                                                                                                                        & 18.07                 & 11.73                 & 2                   & 67                 \\
Travel time (mins): auto (fall   2021)       &                                                                                                                                                                                                                                                                                        & 19.31                 & 14.32                 & 2                   & 92                 \\\hline\\[-1.8ex]
PT usage   frequency (pre-pandemic)          & \multirow{8}{*}{\begin{tabular}[c]{@{}l@{}}Frequency of usage regarding different modes   (recoded\\      as monthly usage frequencies$^1$: never = 0, rarely = 1, a\\      few times a month = 3, once a week = 4, several times a\\      week = 16 (4*4), daily = 28 (4*7)\end{tabular}} & 9.23                  & 10.89                 & 0                   & 28                 \\
PT usage frequency (fall 2021)               &                                                                                                                                                                                                                                                                                        & 6.52                  & 9.50                  & 0                   & 28                 \\
Car usage frequency   (pre-pandemic)         &                                                                                                                                                                                                                                                                                        & 17.12                 & 11.77                 & 0                   & 28                 \\
Car usage frequency (fall   2021)            &                                                                                                                                                                                                                                                                                        & 17.08                 & 11.56                 & 0                   & 28                 \\
TNC usage frequency   (pre-pandemic)         &                                                                                                                                                                                                                                                                                        & 3.96                  & 6.56                  & 0                   & 28                 \\
TNC usage frequency (fall   2021)            &                                                                                                                                                                                                                                                                                        & 3.15                  & 5.90                  & 0                   & 28                 \\
Walk frequency (pre-pandemic)                &                                                                                                                                                                                                                                                                                        & 11.15                 & 11.32                 & 0                   & 28                 \\
Walk frequency (fall 2021)                   &                                                                                                                                                                                                                                                                                        & 11.39                 & 11.33                 & 0                   & 28                 \\\hline\\[-1.8ex]
\multicolumn{6}{l}{\textit{\textbf{2. Travel   attitudinal variables}}}                                                                                                                                                                                                                                                                                                                                                          \\\\[-1.8ex]
Risk perception: car                         & \multirow{6}{*}{\begin{tabular}[c]{@{}l@{}}Perceived risk of becoming infected with COVID-19\\      while using different travel modes (measured on a 5-\\      point Likert scale ranging from extremely unlikely to\\      extremely likely)\end{tabular}}                           & 0.85                  & 1.29                  & 0                   & 4                  \\
Risk perception: TNC                         &                                                                                                                                                                                                                                                                                        & 1.93                  & 1.06                  & 0                   & 4                  \\
Risk perception: bus                         &                                                                                                                                                                                                                                                                                        & 2.44                  & 1.07                  & 0                   & 4                  \\
Risk perception: subway                      &                                                                                                                                                                                                                                                                                        & 2.48                  & 1.09                  & 0                   & 4                  \\
Risk perception: commuter rail               &                                                                                                                                                                                                                                                                                        & 2.36                  & 1.09                  & 0                   & 4                  \\
Risk perception: walking                     &                                                                                                                                                                                                                                                                                        & 0.79                  & 1.11                  & 0                   & 4                  \\\hline\\[-1.8ex]
\multicolumn{6}{l}{\textit{\textbf{3. WFH-related   variables}}}                                                                                                                                                                                                                                                                                                                                                                 \\\\[-1.8ex]
On-site working days                         & Number of days working at the job location in a   week                                                                                                                                                                                                                                 & 2.21                  & 2.23                  & 0                   & 7                  \\
Total working days                           & Number of total working days                                                                                                                                                                                                                                                           & 4.32                  & 1.84                  & 0                   & 7                  \\
\multirow{2}{*}{On-site requirement}         & “Does your job have to be conducted at the job                                                                                                                                                                                                                                         & \multirow{2}{*}{0.34} & \multirow{2}{*}{0.47} & \multirow{2}{*}{0}  & \multirow{2}{*}{1} \\
                                             & location?” (yes = 1; no = 0)                                                                                                                                                                                                                                                           &                       &                       &                     &                    \\
\multirow{2}{*}{Reimbursement}               & “Does your employer reimburse you for any portion of                                                                                                                                                                                                                                   & \multirow{2}{*}{0.22} & \multirow{2}{*}{0.41} & \multirow{2}{*}{0}  & \multirow{2}{*}{1} \\
                                             & the cost of remote work?” (yes = 1; no = 0)                                                                                                                                                                                                                                            &                       &                       &                     &                    \\\hline\\[-1.8ex]
\multicolumn{6}{l}{\textit{\textbf{Evaluation of WFH in   comparison to in-person work for the following aspects (much worse than   in-person work = -2;}}}                                                                                                                                                                                                                                                                      \\
\multicolumn{6}{l}{\textit{\textbf{worse than in-person   work = -1; about the same = 0; better than in-person work = 1; much better   than in-person work = 2)}}}                                                                                                                                                                                                                                                               \\\\[-1.8ex]
WFH satisfaction: productivity               & Work productivity                                                                                                                                                                                                                                                                      & 0.07                  & 1.03                  & -2                  & 2                  \\
WFH satisfaction: focus                      & Focus on work                                                                                                                                                                                                                                                                          & 0.02                  & 1.10                  & -2                  & 2                  \\
WFH satisfaction: creativity                 & Creativity at work                                                                                                                                                                                                                                                                     & 0.23                  & 1.00                  & -2                  & 2                  \\
WFH satisfaction: environment                & Quality of the work environment                                                                                                                                                                                                                                                        & 0.37                  & 1.06                  & -2                  & 2                  \\
WFH satisfaction:   collaboration            & Collaboration with my colleagues                                                                                                                                                                                                                                                       & -0.43                 & 0.99                  & -2                  & 2                  \\
WFH satisfaction: feeling   connected        & Feeling connected to my colleagues                                                                                                                                                                                                                                                     & -0.54                 & 1.04                  & -2                  & 2                  \\
WFH satisfaction: work-life   balance        & Work-life balance                                                                                                                                                                                                                                                                      & 0.56                  & 1.15                  & -2                  & 2                  \\
Overall WFH satisfaction                     & Overall satisfaction                                                                                                                                                                                                                                                                   & 0.41                  & 1.01                  & -2                  & 2                  \\

\midrule
\multicolumn{6}{l}{\textit{Note:} $^1$We take the mean of each frequency bracket in the modeling process.}  
\end{tabular}}
\end{table}
\FloatBarrier
\FloatBarrier
\begin{table}[h]
  \caption{(Cont.) Descriptive statistics}
  \label{tab_desc2} 
\resizebox{\textwidth}{!}{%
\begin{tabular}{@{}llllll@{}}
\toprule
Variable                                     & Description                                                                                                                                                                                                                                                                  & Mean   & Std.dev & Min  & Max    \\ \midrule
\multicolumn{6}{l}{\textit{\textbf{4. Socio-demographic   variables}}}\\\\[-1.8ex]
Car ownership (fall 2021)                    & \multirow{2}{*}{ Number of   cars owned by the household} & 1.47                  & 0.88                  & 0                   & 3                  \\
Car ownership (pre-pandemic)  &                                                                                                                                                                                                                                                                                        & 1.44                  & 0.88                  & 0                   & 3                  \\\hline\\[-1.8ex]
\multirow{3}{*}{Change in car ownership}     & Difference of the number of cars owned by the                                                                                                                                                                                                                                          & \multirow{3}{*}{0.03} & \multirow{3}{*}{0.44} & \multirow{3}{*}{-2} & \multirow{3}{*}{2} \\
                                             & household in the fall of 2021 and that in   the                                                                                                                                                                                                                                 &                       &                       &                     &                    \\
                                             & pre-pandemic period                                                                                                                                                                                                                                                                    &                       &                       &                     &                    \\\hline\\[-1.8ex]
Age                                          & Age (years)                                                                                                                                                                                                                                                                            & 34.68                 & 13.15                 & 18                  & 79.5               \\
Female                                       & Gender (female = 1; male = 0)                                                                                                                                                                                                                                                          & 0.48                  & 0.50                  & 0                   & 1                  \\
Ethnicity: black or African   American       & Ethnicity (black or African American = 1; other =   0)                                     & 0.27                  & 0.45                  & 0                   & 1                  \\
Ethnicity: white       & Ethnicity (white = 1; other =   0)                                     & 0.56                  & 0.50                  & 0                   & 1                  \\
Ethnicity: Asian       & Ethnicity (Asian = 1; other =   0)                                     & 0.11                  & 0.32                  & 0                   & 1                  \\
\multirow{2}{*}{Education level}             & Highest education level 0-5 ranging from “less than   high                                                                                                                                                                                                                             & \multirow{2}{*}{3.32} & \multirow{2}{*}{1.36} & \multirow{2}{*}{0}  & \multirow{2}{*}{5} \\
                                             & school diploma” to “graduate or professional   degree”                                                                                                                                                                                                                                 &                       &                       &                     &                    \\
Household size                               & Number of people in the household                                                                                                                                                                                                                                                      & 2.66                  & 1.25                  & 1                   & 5                  \\
Annual household income ($k$)                 & Household annual income (in 1,000)                                                                                                                                                                                                                                                    & 86.77                 & 59.52                 & 5                   & 225                \\\hline\\[-1.8ex]
Job: Community and personal   services       & \multirow{9}{*}{Occupation (0/1)}                                                                                                                                                                                                                                                      & 0.03                  & 0.17                  & 0                   & 1                  \\
Job: Healthcare                              &                                                                                                                                                                                                                                                                                        & 0.10                  & 0.31                  & 0                   & 1                  \\
Job: Labourers                               &                                                                                                                                                                                                                                                                                        & 0.10                  & 0.30                  & 0                   & 1                  \\
Job: Managers                                &                                                                                                                                                                                                                                                                                        & 0.04                  & 0.18                  & 0                   & 1                  \\
Job: Others                                  &                                                                                                                                                                                                                                                                                        & 0.05                  & 0.22                  & 0                   & 1                  \\
Job: Professionals                           &                                                                                                                                                                                                                                                                                        & 0.32                  & 0.47                  & 0                   & 1                  \\
Job: Sales                                   &                                                                                                                                                                                                                                                                                        & 0.06                  & 0.24                  & 0                   & 1                  \\
Job: Student                                 &                                                                                                                                                                                                                                                                                        & 0.08                  & 0.28                  & 0                   & 1                  \\
Job: Technicians                             &                                                                                                                                                                                                                                                                                        & 0.10                  & 0.31                  & 0                   & 1                  \\\hline\\[-1.8ex]
\multicolumn{6}{l}{\textit{\textbf{5. Built environment   variables}}}                                                                                                                                                                                                                                                                                                                                                           \\\\[-1.8ex]
Distance to CBD (pre-pandemic)               & Distance of the home location to the CBD (Boston City                                                                                                                                                                                                                                  & 18.72                 & 18.39                 & 0.12          & 74.24        \\
Distance to CBD (fall 2021)                  & Hall) in kilometer                                                                                                                                                                                                                                                                     & 19.21                 & 18.57                 & 0.13           & 74.51         \\ \bottomrule
\end{tabular}}
\end{table}
\FloatBarrier

\noindent 

\subsection{4.3 Analytical approach}
This study tries to understand the impact of COVID-19 on WFH, frequency of use for various transportation modes as well as commuting mode choices in the fall of 2021. As such, three separate approaches were taken. The first estimates a travel mode frequency prediction model, the second identifies factors that influence the number of on-site working days and the overall WFH satisfaction. The third estimates a commuting mode choice prediction model.\\

\subsubsection{4.3.1 Travel mode usage prediction}
\begin{figure}[!t]
    \centering
    \includegraphics[width=6 in]{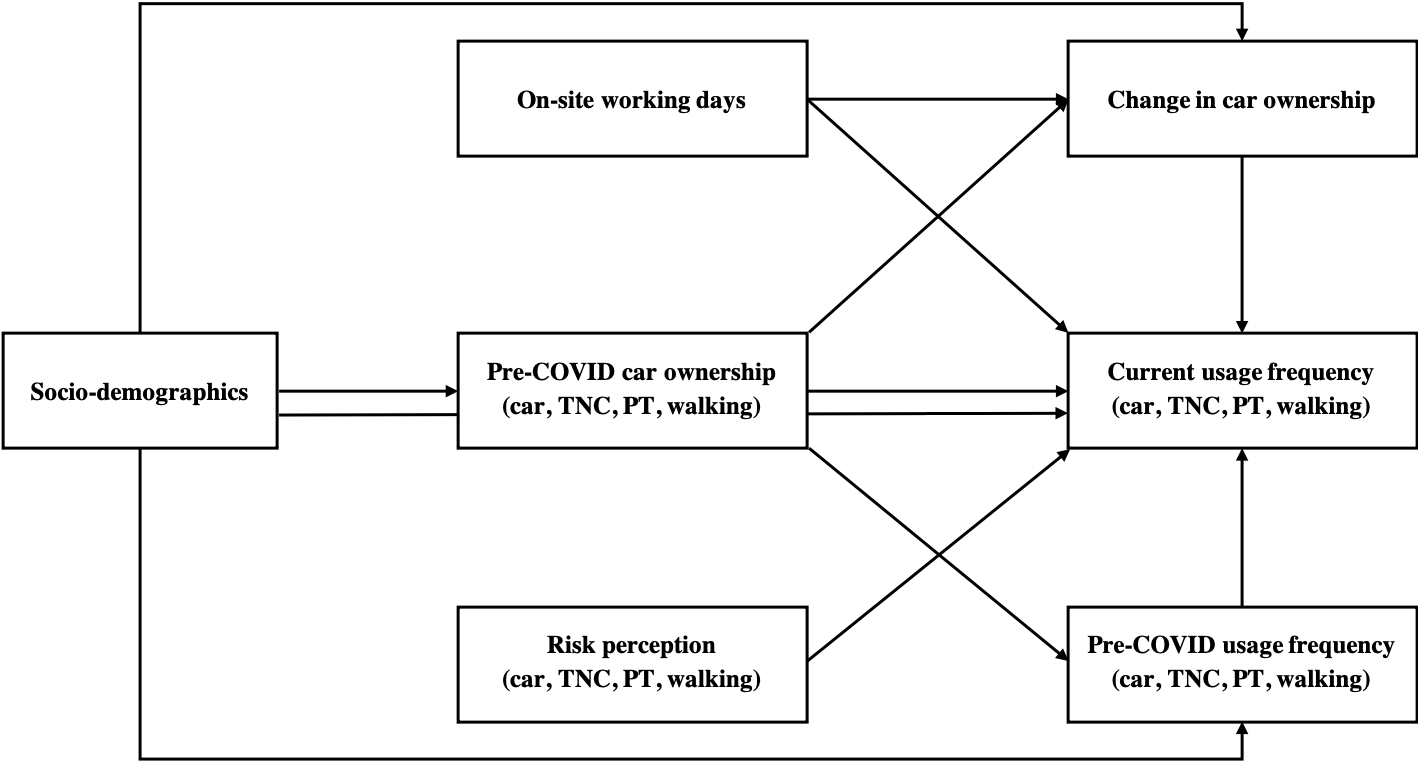}
    \caption{Hypothesised SEM framework for the relationship between the usage frequencies of different travel modes, change in car ownership and WFH}
    \label{fig:sem_structure}
\end{figure}

Given the research gap highlighted in Section \ref{sec:literature}, we use SEM, a multivariate data analysis technique to study the simultaneous influence of travel mode usage frequency, car ownership changes, the number of in-person work days and risk perceptions towards different travel modes in the fall of 2021. The objective of SEM is to estimate the dependency relationship between variables, demonstrating how one variable influences the others and verifying the validity of the model proposed to evaluate these relationships \cite{hair2009multivariate,hair2021primer}. The conceptual framework of the SEM model is illustrated in Figure \ref{fig:sem_structure}. \\

\noindent Several hypotheses are tested. First, the change in car ownership compared to the pre-COVID period is expected to influence the frequency of car and non-car travel modes in the fall of 2021. A previous study has shown that, in the U.S., the perceived value of owning and using a car increased by 260\% during the pandemic when compared with the pre-pandemic period \cite{moody2021value}. As such, we want to explore how the change in car ownership would affect the frequency of use for various travel modes.\\


\noindent Second, the number of in-person working days is expected to be positively correlated with the usage frequencies of various travel modes. Surveys conducted during the early days of the COVID-19 pandemic in Australia showed that a higher number of days of WFH is associated with a lower number of one-way weekly commuter trips \cite{hensher2022impact}, and we want to test whether the total frequency (for all trips, not just commuting trips) is affected. In this research, we aim to not only test the existence of the correlation between the number of in-person working days and the usage frequencies of various travel modes in the fall of 2021, but also study how this relationship varies across different travel modes. \\

\noindent Third, we hypothesize that perceived risks of contracting COVID-19 when using different travel modes influence the number of on-site working days, the change in car ownership and the usage of different travel modes in the fall of 2021. We also hypothesize that the number of in-person working days should be positively correlated with an increased probability of car ownership, since people are more likely to buy cars if they commute often.\\ 

\noindent We adopt an SEM to test the hypothesized relationship using the $Lavaan$ package in R Studio. One key advantage of the SEM method is the flexibility to simultaneously capture the direct and indirect effects among variables. It also allows us to compare the magnitudes of the effects across different variables. For instance, we can identify which travel modes have been most affected by increases in car ownership.

\subsubsection{4.3.2 WFH predictions}
\begin{figure}[!t]
    \centering
    \includegraphics[width=5 in]{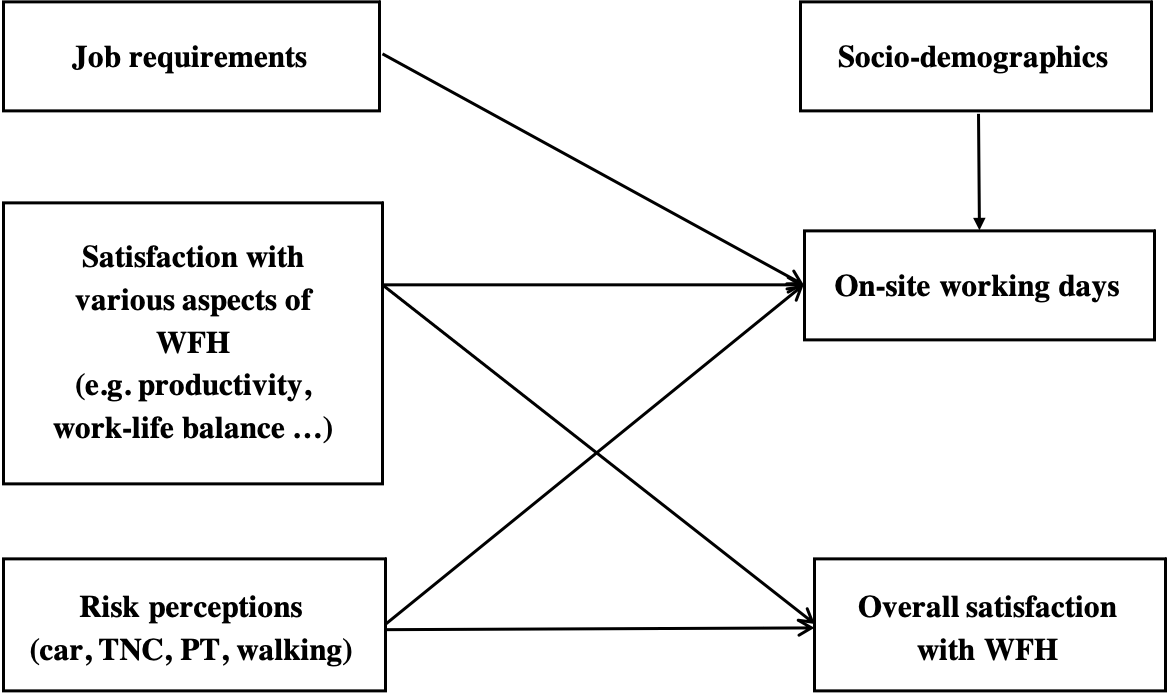}
    \caption{Hypothesised SEM framework for the predictions of on-site working days and WFH satisfaction}
    \label{fig:sem_structure_wfh}
\end{figure}

To understand what factors influence respondents' perceptions and behavior related to WFH, we build another SEM model to estimate the number of on-site working days and overall satisfaction towards WFH. The conceptual framework is presented in Figure \ref{fig:sem_structure_wfh} which shows several hypotheses tested in the model.
First, job requirements and attitudes towards WFH are expected to influence the number of in-person working days. This hypothesis is grounded in the Theory of Planned Behavior, which states that strong favorable attitudes and greater perceived behavioral control (\textit{i.e.} the perceived practical ease or difficulty of performing the behavior) are usually associated with a strong intention to behave in a particular fashion \cite{ajzen1991theory,jain2022covid}. Various aspects of attitudes towards WFH (e.g. work productivity, work environment ...) are expected to affect the number of in-person working days differently, and it is essential to identify the relative roles of these various attitudes in terms of explaining the number of in-person working days, as it can help employers understand what drives people to choose on-site working over WFH. Besides, we hypothesize that people's perceived risk of becoming infected with COVID-19 while using different travel modes may affect their on-site working days. \\

\noindent Second, we also expect that the overall satisfaction towards WFH is affected differently by different dimensions of attitudes towards WFH. The perceived risk of different travel modes may also affect people's overall satisfaction towards WFH. In addition, we hypothesize that the number of on-site working days correlates with overall WFH satisfaction.

\subsubsection{4.3.3 Commuting mode choice prediction}

\noindent While the previous section focuses on predicting the frequency of use for various travel modes, in this section we aim at predicting respondents' primary commuting mode to work/school. Discrete choice analysis is the modeling of choice from a set of exclusive and collectively exhaustive alternatives \cite{ben1985discrete,zheng2021equality}. For this problem, we are trying to predict the respondents' primary commuting mode choices among several discrete alternatives, thus a discrete choice modeling approach is adopted. Only the respondents who are employed/self-employed or students are included in this part of analysis. \\

\noindent The primary commuting mode model contains four alternatives: 1) auto (including car driver, car passenger and taxi or ride hailing); 2) public transport (including bus, subway and commuter rail); 3) active modes (including walking and biking); 4) WFH (no commute). In the survey, we asked not only the respondents' primary commuting modes for the time the survey was conducted, but also their primary commuting modes before COVID-19. With this information, we are able to conduct the same discrete choice analysis for the pre-COVID and the fall of 2021 separately and compare the results. For samples that choose WFH in each period, we exclude those indicating that the origin and destination of a typical commuting trip once the COVID-19 ceases to be a pandemic are the same, because in that case it is impossible for the respondents to choose travel modes other than WFH. This leaves us with only 3 valid samples that chose WFH for the pre-pandemic time. Therefore, for the pre-pandemic period, we excluded responses that chose WFH and only estimated for a choice set with three alternatives: active, auto and transit modes. After data cleaning, there were 349 valid samples for the pre-pandemic period and 335 valid samples for the fall of 2021.\\

\noindent In the discrete choice models, the utility of working from home is given by:

{\setlength{\mathindent}{0pt}
  \begin{equation}
\begin{aligned}
U_{WFH} = &ACS_{WFH}+\sum_{q}^{}\beta_{WFH,q} \cdot z_{q}+\sum_{d}^{}\beta_{WFH,d} \cdot x_{d}+\beta_{WFH,r}\cdot reimburse+\beta_{WFH,s}\cdot satisfaction+\varepsilon_{WFH}
  \end{aligned}
  \label{eq_wfh_util}
\end{equation}}


\noindent where $z_{q}$ represents the socio-demographic variables including gender, race, income, education level, occupation type, car ownership and household size. $x_{d}$ represents the built environment variables including the distance of the individual's home to the CBD (in this case Boston City Hall). $reimburse$ indicates whether the employer reimbursed the individual for any portion of WFH cost. $satisfaction$ indicates the individual's general satisfaction towards WFH as measured on a 5-point Likert scale. Note that the WFH alternative was not included in the pre-pandemic period. \\

\noindent The utility functions for the modal alternatives are given by:
{\setlength{\mathindent}{0pt}
  \begin{equation}
\begin{aligned}
U_{active} = &ACS_{active}+\beta_{active,TT}\cdot TT_{active}+\varepsilon_{active}\\
U_{auto} = &ACS_{auto}+\beta_{auto,TT}\cdot TT_{auto}+\sum_{q}^{}\beta_{auto,q} \cdot z_{q}+\sum_{d}^{}\beta_{auto,d} \cdot x_{d}+\varepsilon_{auto}\\
U_{transit} = &ACS_{transit}+\beta_{transit,TT}\cdot TT_{transit}+\sum_{q}^{}\beta_{transit,q} \cdot z_{q}+\sum_{d}^{}\beta_{transit,d} \cdot x_{d}+\varepsilon_{transit}
   \end{aligned}
   \label{eq_mode_util}
\end{equation}}


\noindent where $TT_{Mode_j}$ represents the travel time for the travel mode $j$, obtained through the Google Maps API. As with the WFH alternative, the socio-demographic variables $z_q$ and the built environment variables $x_d$ are included in the utility function.

\section{5. Results}
\label{sec:results}

\subsection{5.1 Travel mode usage prediction}

\begin{figure}[H]
    \centering
    \begin{minipage}{\textwidth} 
\includegraphics[width=\textwidth]{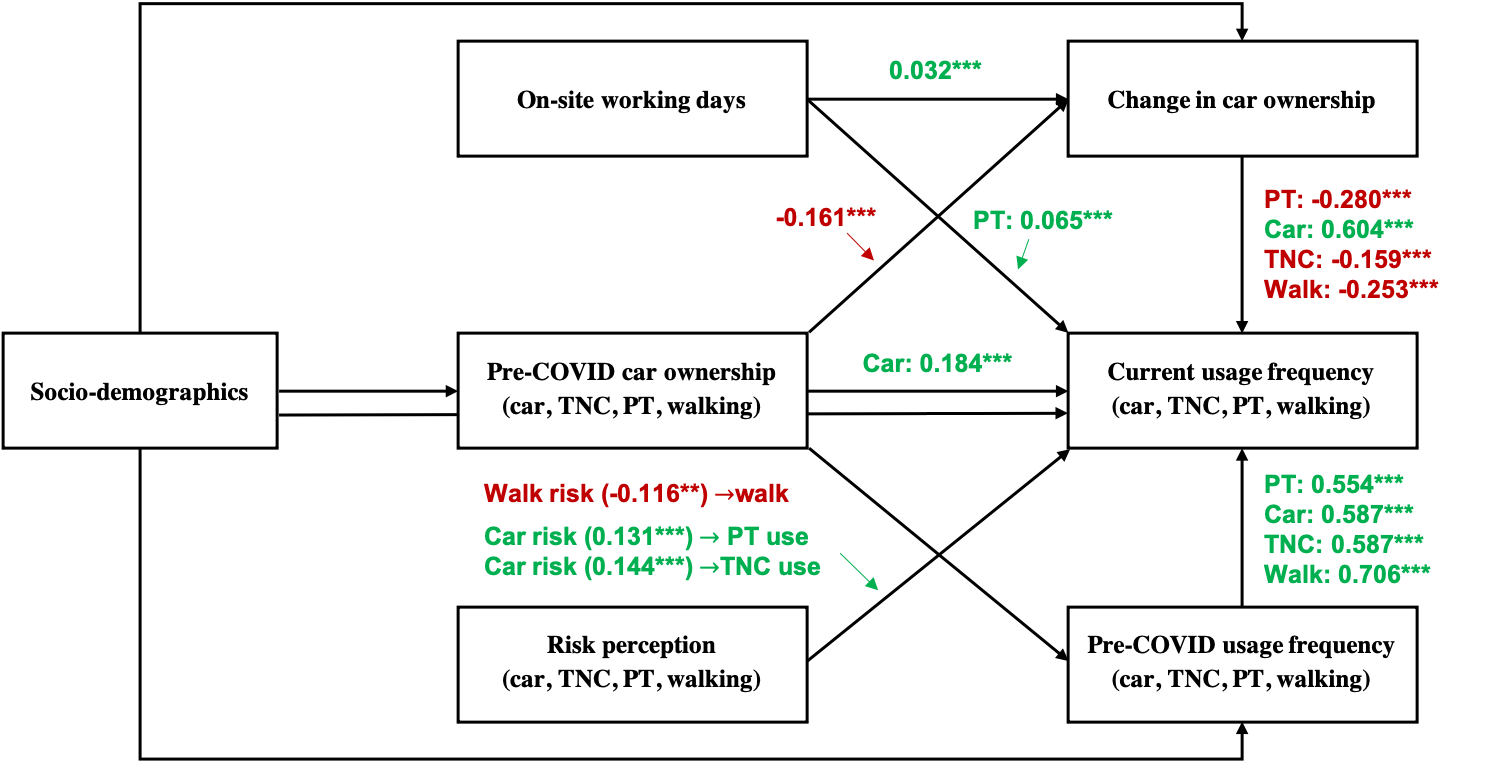}
{\footnotesize \par Note: green indicates positive coefficients, whereas red indicates negative coefficients. ``Current usage frequency'' refers to the usage frequency in the fall of 2021. \par}
\end{minipage}
    
    \caption{Travel mode frequency prediction result}
    \label{fig:sem_output_freq}
\end{figure}

The results for the usage frequencies prediction are reported in Table \ref{result_freq1} (for public transit and car) and Table \ref{result_freq2} (for TNC and walking), whereas the results for predicting change in car ownership is reported in Table \ref{result_carown}. We put the coefficients of the key variables in Figure \ref{fig:sem_output_freq} for clarity.\\

\noindent To evaluate the model performance, two indices are used to measure the goodness fit of the overall model: Comparative Fit Index (CFI) and the Root Mean Square Error of Approximation (RMSEA). As suggested by \citet{ullman2012structural}, CFI values greater than 0.95 and RMSEA values of 0.06 or less are often indicative of a good fitting model. We find that the CFI values for transit, car, TNC, walking and WFH frequency predictions are 0.96, 0.96, 0.95, 0.96 and 1.00. The RMSEA for transit, car, TNC, walking and WFH frequency predictions are 0.05, 0.05, 0.06, 0.06 and 0.00. These values demonstrate good fits for all of the models.



\subsubsection{5.1.1 Travel mode frequency prediction}
\noindent First, looking at the prediction results for the usage of different travel modes in the fall of 2021, we find that unsurprisingly, the pre-COVID usage frequency of a travel mode is the strongest predictor for the usage frequency of that travel mode in the fall of 2021. However, there is still a great variation across the four travel modes. The correlation between the pre-COVID usage frequency and the usage frequency in the fall of 2021 is relatively strong for walking (coeff = 0.706, p$<$0.01), and is relatively weak for PT (coeff = 0.554, p$<$0.01). Specifically, a 1\% increase in pre-COVID walking frequency can explain 0.706\% of the increase of the walking frequency in the fall of 2021, whereas 1\% increase of pre-COVID PT usage frequency can only explain 0.554\% of the increase of the PT usage frequency in the fall of 2021. This result reflects the sluggishness of PT ridership in their rebound, recovering much more slowly than walking. In terms of the car usage, we find that 1\% increase of pre-COVID car usage frequency and car ownership can explain 0.587\% and 0.184\% of the increase of the car usage frequency in the fall of 2021. While a lower level of car ownership prior to the pandemic is typically linked with a higher frequency of other travel modes, we find that this correlation is not significant in the fall of 2021 in our study. In other words, there isn't a strong relationship between pre-pandemic car ownership and the frequency of using other modes of transportation during this specific time period.\\

\noindent While pre-pandemic car ownership only appears to have an impact on car usage in the fall of 2021, we find that change in car ownership in the fall of 2021 compared with the pre-pandemic time significantly affect the usage frequencies of all travel modes.  Growth in car ownership is associated with an increase in the use of cars, and a decrease in the use of PT, TNC and walking in the fall of 2021. When comparing the effects across different modes, we observe that with each additional car owned, the reduction in the frequency of public transportation usage is the largest, at an average of 28\%. The reduction in TNC usage frequency is 15.9\%, and the reduction in walking frequency is 25.3\%.  \\


\noindent The results also show a significant effect of the number of on-site working days on the usage frequency of PT in the fall of 2021. Specifically, one additional on-site working day is associated with a 6.5\% of increase in PT usage frequency. However, this effect is tiny compared with the effect of change in car ownership on PT usage, which indicates that, while there have been proposals for transit agencies to partner with employers and local businesses to plan for a widespread return to the office, the consequent magnitude of the effect may be marginal compared with the effect of people buying new cars during the pandemic. \\

\noindent In terms of the influence of COVID infection risk on travel modes use, the results show that a higher perceived risk of walking reduces the walking frequencies, whereas a lower perceived risk of car reduces PT and TNC usage in the fall of 2021. These results confirm that the overall fall in PT and TNC use compared with the pre-COVID period is partially attributable to people's generally low perceived risk of using cars. \\

\subsubsection{5.1.2 Change in car ownership prediction}

\noindent As shown in Table \ref{result_carown}, the change in car ownership is likely to be higher among individuals whose jobs are healthcare (coeff = 0.262, p$<$0.01), professionals (coeff = 0.220, p$<$0.01), sales (coeff = 0.204, p$<$0.05) or technicians (coeff = 0.218, p$<$0.01). The association between healthcare and change in car ownership is the strongest among all the job categories. Healthcare workers were considered ``essential workers'' during the pandemic, as their duties needed to be conducted on-site, and thus were at high risk for exposure to COVID-19.
It is possible that healthcare workers became car-dependent during the pandemic due to PT service cuts and limited TNC availability. Interestingly, we find no significant relationship between the healthcare job and transportation mode usage. One possible explanation is that the higher degree of car usage among healthcare workers can be attributed to the growth in car ownership among these workers. Another possibility is that the healthcare workers worked continuously in hospitals at a stretch without going home, but more inductive or qualitative analysis is needed to confirm this hypothesis. In terms of the spatial variation, for every 20 km increase in distance from the CBD, our result indicates a corresponding increase in car ownership of only 0.1 units, suggesting a weak relationship between distance to the CBD and car ownership (coeff = 0.005, p$<$0.01). We find no significant correlation between risk perceptions towards various travel modes and change in car ownership.


\subsubsection{5.1.3 The effects of socio-demographic factors}
To understand how the attitudinal and behavioral results vary across different demographics, we highlight some key findings with regards to socio-demographic factors. First, we observe that household size positively correlates with change in car ownership, although the correlation is weak. Specifically, when the household size increases by one, the change in car ownership increases by 0.05 units (coeff=0.053, p$<$0.01). The interpretation of this result is that larger households are in greater need of vehicles for transportation of household members compared with smaller households. \\

\noindent Our results also show evidence of occupational variation in the usage frequencies of different travel modes in the fall of 2021. Specifically, compared with other occupations, people who work in the community and personal services are associated with a 0.49\% higher PT usage frequency (coeff=0.488, p$<$0.1); students are associated with a 0.74\% lower car usage frequency (coeff=-0.735, p$<$0.01) and a 0.6\% higher walking frequency (coeff=0.6, p$<$0.01) 
Technicians on average have a 0.31\% higher car usage frequency (coeff=0.306, p$<$0.1) whereas managers on average are associated with a 0.52\% higher TNC usage frequency (coeff=0.519, p$<$0.01). \\

\subsubsection{5.1.4 The effects of built environment variables}
People living in different parts of Greater Boston may have very different behavior and attitudes towards travel mode usages and WFH, thus we are interested in investigating the effects of the built environment variable (\textit{i.e.} distance to the CBD). Our analysis indicates that distance to the CBD has a modest but statistically significant effect on transportation usage. Tables \ref{result_freq1} and \ref{result_freq2} show that people who live closer to Boston city center tend to use PT and TNC more frequently, as well as walk more often. This relationship likely reflects the fact that the city center has better transportation infrastructure and pedestrian amenities \cite{zheng2023impacts}. Specifically, when the distance to CBD increases by 10 km, the usage frequencies of PT, TNC, and walking reduce by 0.1\%, 0.03\%, and 0.06\%.\\

\noindent In terms of the car usage and ownership, Tables \ref{result_freq1} and \ref{result_carown} show that the distance to CBD has a small but significantly positive association with both the car usage frequency and change in car ownership in the fall of 2021. Specifically,  for every 10 km increase in distance from the CBD, our result indicates a corresponding increase in car usage of 0.05\% and an increase in car ownership of 0.02 unit. These findings suggest that in contrast to the results for walking as well as PT and TNC usage frequencies, people who live in the peripheral area of Greater Boston are more likely to use cars in the fall of 2021, and are also more likely to buy new cars during the pandemic. \\


\FloatBarrier
\begin{table}[h]
  \caption{SEM prediction results: public transit and car usage frequencies predictions}
  \label{result_freq1} 
\resizebox{\textwidth}{!}{%
\begin{tabular}{@{}lllllll@{}}
\toprule
Regression                                   & Coefficient       & R$^2$   &  & Regression                                  & Coefficient       & R$^2$   \\ \midrule
\textbf{PT usage frequency   (pre-pandemic)} &                   & 0.155                &  & \textbf{Car usage frequency (pre-pandemic)} &                   & 0.215                \\
(Intercept)                                  & 1.719 (0.419)***  &                      &  & (Intercept)                                 & 1.830 (0.362)***  &                      \\
Car ownership (pre-pandemic)                 & -0.324 (0.081)*** &                      &  & Car ownership (pre-pandemic)                & 0.520 (0.070)***  &                      \\
Age                                          & -0.017 (0.005)*** &                      &  & Age                                         & -0.005 (0.004)    &                      \\
Income                                       & 0.008 (0.011)     &                      &  & Income                                      & -0.003 (0.010)    &                      \\
Education level                              & 0.048 (0.052)     &                      &  & Education level                             & 0.025 (0.045)     &                      \\
Household size                               & 0.108 (0.056)*    &                      &  & Household size                              & -0.093 (0.048)*   &                      \\
Female                                       & 0.096 (0.128)     &                      &  & Female                                      & -0.094 (0.110)    &                      \\
Ethnicity: white                             & 0.351 (0.267)     &                      &  & Ethnicity: white                            & 0.040 (0.231)     &                      \\
Ethnicity: black or African American         & 0.357 (0.281)     &                      &  & Ethnicity: black or African American        & 0.022 (0.244)     &                      \\
Ethnicity: Asian                             & 0.462 (0.311)     &                      &  & Ethnicity: Asian                            & -0.167 (0.269)    &                      \\
Job: Community and personal services         & -0.019 (0.367)    &                      &  & Job: Community and personal services        & 0.676 (0.317)**   &                      \\
Job: Healthcare                              & 0.116 (0.250)     &                      &  & Job: Healthcare                             & 0.308 (0.217)     &                      \\
Job: Labourers                               & 0.101 (0.239)     &                      &  & Job: Labourers                              & 0.387 (0.206)*    &                      \\
Job: Managers                                & 0.752 (0.350)**   &                      &  & Job: Managers                               & 0.005 (0.303)     &                      \\
Job: Professionals                           & 0.296 (0.199)     &                      &  & Job: Professionals                          & 0.117 (0.172)     &                      \\
Job: Sales                                   & 0.058 (0.284)     &                      &  & Job: Sales                                  & 0.235 (0.246)     &                      \\
Job: Student                                 & 0.385 (0.271)     &                      &  & Job: Student                                & -0.610 (0.234)*** &                      \\
Job: Technicians                             & 0.201 (0.245)     &                      &  & Job: Technicians                            & 0.146 (0.212)     &                      \\
Distance to CBD (pre-pandemic)               & -0.011 (0.003)*** &                      &  & Distance to CBD (pre-pandemic)              & 0.009 (0.003)***  &                      \\ \midrule
\textbf{PT usage frequency (fall 2021)}      &                   & 0.480                &  & \textbf{Car usage frequency (fall 2021)}    &                   & 0.542                \\
(Intercept)                                  & 0.733 (0.345)**   &                      &  & (Intercept)                                 & 0.949 (0.317)***  &                      \\
PT usage frequency (pre-pandemic)            & 0.554 (0.035)***  &                      &  & Car usage frequency (pre-pandemic)          & 0.587 (0.037)***  &                      \\
Car ownership (pre-pandemic)                 & -0.028 (0.062)    &                      &  & Car ownership (pre-pandemic)                & 0.184 (0.059)***  &                      \\
Change in car ownership                      & -0.280 (0.105)*** &                      &  & Change in car ownership                     & 0.604 (0.096)***  &                      \\
Risk perception: car                         & 0.131 (0.044)***  &                      &  & Risk perception: car                        & -0.008 (0.040)    &                      \\
Risk perception: TNC                         & 0.049 (0.057)     &                      &  & Risk perception: TNC                        & -0.052 (0.052)    &                      \\
Risk perception: PT                          & -0.050 (0.058)    &                      &  & Risk perception: PT                         & 0.015 (0.053)     &                      \\
Risk perception: walking                     & -0.077 (0.050)    &                      &  & Risk perception: walking                    & 0.019 (0.046)     &                      \\
On-site working days                         & 0.065 (0.021)***  &                      &  & On-site working days                        & 0.012 (0.019)     &                      \\
Age                                          & 0.005 (0.004)     &                      &  & Age                                         & -0.002 (0.003)    &                      \\
Income                                       & -0.004 (0.008)    &                      &  & Income                                      & -0.010 (0.008)    &                      \\
Education level                              & -0.080 (0.039)**  &                      &  & Education level                             & -0.016 (0.035)    &                      \\
Household size                               & -0.046 (0.042)    &                      &  & Household size                              & 0.017 (0.038)     &                      \\
Female                                       & 0.072 (0.095)     &                      &  & Female                                      & -0.026 (0.087)    &                      \\
Ethnicity: white                             & -0.100 (0.197)    &                      &  & Ethnicity: white                            & -0.129 (0.179)    &                      \\
Ethnicity: black or African American         & -0.085 (0.209)    &                      &  & Ethnicity: black or African American        & -0.142 (0.190)    &                      \\
Ethnicity: Asian                             & -0.339 (0.229)    &                      &  & Ethnicity: Asian                            & -0.268 (0.208)    &                      \\
Job: Community and personal services         & 0.488 (0.270)*    &                      &  & Job: Community and personal services        & 0.230 (0.247)     &                      \\
Job: Healthcare                              & 0.006 (0.190)     &                      &  & Job: Healthcare                             & 0.144 (0.173)     &                      \\
Job: Labourers                               & 0.223 (0.181)     &                      &  & Job: Labourers                              & 0.121 (0.166)     &                      \\
Job: Managers                                & -0.096 (0.258)    &                      &  & Job: Managers                               & -0.060 (0.234)    &                      \\
Job: Professionals                           & -0.005 (0.150)    &                      &  & Job: Professionals                          & 0.107 (0.137)     &                      \\
Job: Sales                                   & -0.098 (0.215)    &                      &  & Job: Sales                                  & 0.173 (0.197)     &                      \\
Job: Student                                 & 0.309 (0.202)     &                      &  & Job: Student                                & -0.735 (0.185)*** &                      \\
Job: Technicians                             & -0.007 (0.182)    &                      &  & Job: Technicians                            & 0.306 (0.166)*    &                      \\
Distance to CBD (fall 2021)                  & -0.010 (0.002)*** &                      &  & Distance to CBD (fall 2021)                 & 0.005 (0.002)**   &                      \\ \bottomrule
\multicolumn{5}{l}{\textit{Note:} Standard errors are reported in parentheses. Dependent variables are denoted in bold. Log-transformation}  & \multicolumn{2}{r}{$^{*}$p$<$0.1; $^{**}$p$<$0.05; $^{***}$p$<$0.01} \\ 
\multicolumn{5}{l}{ has been adopted to the usage frequency variables for all the modes. ``Risk perception: PT'' is computed as the }  & \multicolumn{2}{r}{} \\ 
\multicolumn{5}{l}{average risk perception towards bus, subway and commuter rail. Only the significant coefficients are reported.}  & \multicolumn{2}{r}{} \\ 
\end{tabular}
} 
\end{table}
\FloatBarrier

\FloatBarrier
\begin{table}[h]
  \caption{SEM prediction results: TNC usage and walk frequencies predictions}
  \label{result_freq2} 
\resizebox{\textwidth}{!}{%
\begin{tabular}{@{}lllllll@{}}
\toprule
Regression                                    & Coefficient       & R$^2$   &  & Regression                             & Coefficient       & R$^2$   \\ \midrule
\textbf{TNC usage frequency   (pre-pandemic)} &                   & 0.175                &  & \textbf{Walk frequency (pre-pandemic)} &                   & 0.114                \\
(Intercept)                                   & 1.746 (0.308)***  &                      &  & (Intercept)                            & 1.972 (0.430)***  &                      \\
Car ownership (pre-pandemic)                  & -0.194 (0.060)*** &                      &  & Car ownership (pre-pandemic)           & -0.308 (0.083)*** &                      \\
Age                                           & -0.017 (0.004)*** &                      &  & Age                                    & -0.003 (0.005)    &                      \\
Income                                        & 0.006 (0.008)     &                      &  & Income                                 & -0.002 (0.011)    &                      \\
Education level                               & 0.013 (0.038)     &                      &  & Education level                        & 0.029 (0.054)     &                      \\
Household size                                & 0.055 (0.041)     &                      &  & Household size                         & 0.077 (0.057)     &                      \\
Female                                        & 0.167 (0.094)*    &                      &  & Female                                 & 0.215 (0.131)     &                      \\
Ethnicity: white                              & 0.043 (0.196)     &                      &  & Ethnicity: white                       & 0.456 (0.274)*    &                      \\
Ethnicity: black or African American          & 0.303 (0.207)     &                      &  & Ethnicity: black or African American   & 0.174 (0.289)     &                      \\
Ethnicity: Asian                              & -0.099 (0.228)    &                      &  & Ethnicity: Asian                       & 0.637 (0.319)**   &                      \\
Job: Community and personal services          & 0.301 (0.269)     &                      &  & Job: Community and personal services   & -0.979 (0.377)*** &                      \\
Job: Healthcare                               & -0.050 (0.184)    &                      &  & Job: Healthcare                        & -0.169 (0.257)    &                      \\
Job: Labourers                                & -0.288 (0.175)    &                      &  & Job: Labourers                         & -0.473 (0.245)*   &                      \\
Job: Managers                                 & -0.214 (0.257)    &                      &  & Job: Managers                          & -0.657 (0.360)*   &                      \\
Job: Professionals                            & -0.048 (0.146)    &                      &  & Job: Professionals                     & -0.237 (0.205)    &                      \\
Job: Sales                                    & -0.005 (0.209)    &                      &  & Job: Sales                             & -0.493 (0.292)*   &                      \\
Job: Student                                  & -0.568 (0.199)*** &                      &  & Job: Student                           & 0.265 (0.278)     &                      \\
Job: Technicians                              & -0.090 (0.180)    &                      &  & Job: Technicians                       & -0.356 (0.252)    &                      \\
Distance to CBD (pre-pandemic)                & -0.007 (0.002)*** &                      &  & Distance to CBD (pre-pandemic)         & -0.007 (0.003)**  &                      \\ \midrule
\textbf{TNC usage frequency (fall   2021)}    &                   & 0.521                &  & \textbf{Walk frequency (fall 2021)}    &                   & 0.619                \\
(Intercept)                                   & 0.281 (0.247)     &                      &  & (Intercept)                            & 0.965 (0.312)***  &                      \\
TNC usage frequency (pre-pandemic)            & 0.587 (0.034)***  &                      &  & Walk frequency (pre-pandemic)          & 0.706 (0.031)***  &                      \\
Car ownership (pre-pandemic)                  & -0.008 (0.044)    &                      &  & Car ownership (pre-pandemic)           & -0.065 (0.056)    &                      \\
Change in car ownership                       & -0.159 (0.074)**  &                      &  & Change in car ownership                & -0.253 (0.095)*** &                      \\
Risk perception: car                          & 0.144 (0.031)***  &                      &  & Risk perception: car                   & 0.033 (0.040)     &                      \\
Risk perception: TNC                          & -0.031 (0.040)    &                      &  & Risk perception: TNC                   & 0.040 (0.051)     &                      \\
Risk perception: PT                           & 0.044 (0.041)     &                      &  & Risk perception: PT                    & 0.029 (0.052)     &                      \\
Risk perception: walking                      & 0.043 (0.035)     &                      &  & Risk perception: walking               & -0.116 (0.045)*** &                      \\
On-site working days                          & 0.019 (0.015)     &                      &  & On-site working days                   & -0.001 (0.019)    &                      \\
Age                                           & 0.000 (0.003)     &                      &  & Age                                    & 0.003 (0.003)     &                      \\
Income                                        & 0.000 (0.006)     &                      &  & Income                                 & 0.004 (0.007)     &                      \\
Education level                               & -0.026 (0.027)    &                      &  & Education level                        & -0.096 (0.035)*** &                      \\
Household size                                & -0.044 (0.029)    &                      &  & Household size                         & -0.008 (0.038)    &                      \\
Female                                        & -0.115 (0.067)*   &                      &  & Female                                 & 0.082 (0.086)     &                      \\
Ethnicity: white                              & -0.017 (0.139)    &                      &  & Ethnicity: white                       & -0.125 (0.178)    &                      \\
Ethnicity: black or African American          & -0.011 (0.147)    &                      &  & Ethnicity: black or African American   & -0.350 (0.188)*   &                      \\
Ethnicity: Asian                              & -0.075 (0.161)    &                      &  & Ethnicity: Asian                       & -0.182 (0.207)    &                      \\
Job: Community and personal services          & 0.246 (0.190)     &                      &  & Job: Community and personal services   & 0.177 (0.245)     &                      \\
Job: Healthcare                               & 0.073 (0.134)     &                      &  & Job: Healthcare                        & 0.164 (0.171)     &                      \\
Job: Labourers                                & 0.073 (0.128)     &                      &  & Job: Labourers                         & 0.074 (0.164)     &                      \\
Job: Managers                                 & 0.519 (0.181)***  &                      &  & Job: Managers                          & 0.115 (0.232)     &                      \\
Job: Professionals                            & 0.090 (0.106)     &                      &  & Job: Professionals                     & 0.046 (0.135)     &                      \\
Job: Sales                                    & 0.097 (0.152)     &                      &  & Job: Sales                             & -0.180 (0.195)    &                      \\
Job: Student                                  & 0.192 (0.143)     &                      &  & Job: Student                           & 0.600 (0.182)***  &                      \\
Job: Technicians                              & 0.032 (0.128)     &                      &  & Job: Technicians                       & 0.166 (0.164)     &                      \\
Distance to CBD (fall 2021)                   & -0.003 (0.002)*   &                      &  & Distance to CBD (fall 2021)            & -0.006 (0.002)*** &                      \\ \bottomrule

\multicolumn{5}{l}{\textit{Note:} Standard errors are reported in parentheses. Dependent variables are denoted in bold. Log-transformation}  & \multicolumn{2}{r}{$^{*}$p$<$0.1; $^{**}$p$<$0.05; $^{***}$p$<$0.01} \\ 
\multicolumn{5}{l}{ has been adopted to the usage frequency variables for all the modes. ``Risk perception: PT'' is computed as the }  & \multicolumn{2}{r}{} \\ 
\multicolumn{5}{l}{average risk perception towards bus, subway and commuter rail. Only the significant coefficients are reported.}  & \multicolumn{2}{r}{} \\ 
\end{tabular}%
} 
\end{table}
\FloatBarrier

\FloatBarrier
\begin{table}[h]
\centering
\caption{SEM prediction results: pre-pandemic car ownership and change in car ownership}
  \label{result_carown} 
\resizebox{\textwidth}{!}{%
\begin{tabular}{@{}lllllll@{}}
\toprule
Regression                            & Coefficient      & R$^2$ &  & Regression                           & Coefficient       & R$^2$ \\ \midrule
\textbf{Car ownership (pre-pandemic)} &                  & 0.137                &  & \textbf{Change in car ownership}     &                   & 0.137                \\
(Intercept)                           & 0.606 (0.249)**  &                      &  & (Intercept)                          & -0.230 (0.157)    &                      \\
Age                                   & -0.004 (0.003)   &                      &  & Car ownership (pre-pandemic)         & -0.161 (0.027)*** &                      \\
Income                                & 0.023 (0.007)*** &                      &  & On-site working days                 & 0.032 (0.010)***  &                      \\
Education level                       & 0.016 (0.031)    &                      &  & Risk perception: car                 & -0.010 (0.020)    &                      \\
Household size                        & 0.284 (0.030)*** &                      &  & Risk perception: TNC                 & 0.030 (0.026)     &                      \\
Female                                & -0.162 (0.076)** &                      &  & Risk perception: PT                  & 0.002 (0.027)     &                      \\
Ethnicity: white                      & -0.037 (0.160)   &                      &  & Risk perception: walking             & 0.007 (0.023)     &                      \\
Ethnicity: black or African American  & -0.368 (0.168)** &                      &  & Age                                  & 0.002 (0.002)     &                      \\
Ethnicity: Asian                      & -0.129 (0.186)   &                      &  & Income                               & 0.005 (0.004)     &                      \\
Job: Community and personal services  & -0.367 (0.219)*  &                      &  & Education level                      & -0.028 (0.018)    &                      \\
Job: Healthcare                       & 0.045 (0.150)    &                      &  & Household size                       & 0.053 (0.019)***  &                      \\
Job: Labourers                        & 0.076 (0.143)    &                      &  & Female                               & 0.078 (0.044)*    &                      \\
Job: Managers                         & 0.038 (0.210)    &                      &  & Ethnicity: white                     & -0.028 (0.091)    &                      \\
Job: Professionals                    & -0.037 (0.119)   &                      &  & Ethnicity: black or African American & -0.008 (0.096)    &                      \\
Job: Sales                            & 0.087 (0.170)    &                      &  & Ethnicity: Asian                     & -0.001 (0.106)    &                      \\
Job: Student                          & 0.313 (0.161)*   &                      &  & Job: Community and personal services & -0.066 (0.125)    &                      \\
Job: Technicians                      & -0.052 (0.147)   &                      &  & Job: Healthcare                      & 0.262 (0.087)***  &                      \\
Distance to CBD (pre-pandemic)        & 0.009 (0.002)*** &                      &  & Job: Labourers                       & 0.050 (0.084)     &                      \\
                                      &                  &   &  & Job: Managers                        & 0.110 (0.119)     &                      \\
                                      &                  &                      &  & Job: Professionals                   & 0.220 (0.069)***  &                      \\
                                      &                  &                      &  & Job: Sales                           & 0.204 (0.099)**   &                      \\
                                      &                  &                      &  & Job: Student                         & 0.078 (0.093)     &                      \\
                                      &                  &                      &  & Job: Technicians                     & 0.218 (0.083)***  &                      \\
                                      &                  &                      &  & Distance to CBD (fall 2021)          & 0.002 (0.001)*    &                      \\ \bottomrule

\multicolumn{5}{l}{\textit{Note:} Standard errors are reported in parentheses. Dependent variables are denoted in bold. Log-transformation}  & \multicolumn{2}{r}{$^{*}$p$<$0.1; $^{**}$p$<$0.05; $^{***}$p$<$0.01} \\ 
\multicolumn{5}{l}{ has been adopted to the usage frequency variables for all the modes. ``Risk perception: PT'' is computed as the }  & \multicolumn{2}{r}{} \\ 
\multicolumn{5}{l}{average risk perception towards bus, subway and commuter rail. Only the significant coefficients are reported.}  & \multicolumn{2}{r}{} \\ 
\end{tabular}%
} 
\end{table}
\FloatBarrier

\FloatBarrier
\begin{table}[h]
  \caption{SEM prediction results: on-site working days, overall WFH satisfaction}
  \label{result_wfh} 
\resizebox{\textwidth}{!}{%
\begin{tabular}{@{}lllllll@{}}
\toprule
Regression                               & Coefficient       & R$^2$   &  & Regression                           & Coefficient       & R$^2$   \\ \midrule
\textbf{On-site working days}            &                   & 0.497                &  & \textbf{Overall WFH satisfaction}    &                   & 0.621                \\
(Intercept)                              & 0.530 (0.608)     &                      &  & (Intercept)                          & 0.315 (0.226)     &                      \\
Total working days                       & 0.296 (0.053)***  &                      &  & Reimbursement                        & 0.017 (0.079)     &                      \\
Total working days * on-site requirement & 0.558 (0.133)***  &                      &  & Risk perception: car                 & -0.026 (0.033)    &                      \\
On-site requirement                      & -0.378 (0.652)    &                      &  & Risk perception: TNC                 & -0.037 (0.041)    &                      \\
Reimbursement                            & -0.901 (0.203)*** &                      &  & Risk perception: PT                  & 0.056 (0.042)     &                      \\
Risk perception: car                     & -0.092 (0.083)    &                      &  & Risk perception: walking             & 0.038 (0.036)     &                      \\
Risk perception: TNC                     & -0.105 (0.104)    &                      &  & Age                                  & -0.005 (0.003)**  &                      \\
Risk perception: PT                      & -0.067 (0.107)    &                      &  & Income                               & 0.007 (0.006)     &                      \\
Risk perception: walking                 & -0.050 (0.092)    &                      &  & Education level                      & 0.011 (0.026)     &                      \\
Age                                      & -0.006 (0.007)    &                      &  & Household size                       & -0.079 (0.030)*** &                      \\
Income                                   & -0.015 (0.015)    &                      &  & Female                               & -0.051 (0.066)    &                      \\
Education level                          & -0.022 (0.068)    &                      &  & Ethnicity: white                     & 0.053 (0.139)     &                      \\
Household size                           & -0.031 (0.075)    &                      &  & Ethnicity: black or African American & 0.027 (0.148)     &                      \\
Female                                   & -0.048 (0.168)    &                      &  & Ethnicity: Asian                     & -0.025 (0.162)    &                      \\
Ethnicity: white                         & 0.695 (0.354)**   &                      &  & Distance to CBD (fall 2021)          & 0.002 (0.002)     &                      \\
Ethnicity: black or African American     & 0.658 (0.375)*    &                      &  & WFH satisfaction: productivity       & 0.169 (0.049)***  &                      \\
Ethnicity: Asian                         & 0.384 (0.411)     &                      &  & WFH satisfaction: focus              & 0.068 (0.046)     &                      \\
Distance to CBD (fall 2021)              & -0.007 (0.004)    &                      &  & WFH satisfaction: creativity         & 0.190 (0.041)***  &                      \\
WFH satisfaction: productivity           & -0.084 (0.126)    &                      &  & WFH satisfaction: environment        & 0.146 (0.039)***  &                      \\
WFH satisfaction: focus                  & 0.054 (0.116)     &                      &  & WFH satisfaction: collaboration      & 0.055 (0.051)     &                      \\
WFH satisfaction: creativity             & 0.004 (0.105)     &                      &  & WFH satisfaction: feeling connected  & 0.026 (0.047)     &                      \\
WFH satisfaction: environment            & 0.102 (0.099)     &                      &  & WFH satisfaction: work-life balance  & 0.330 (0.032)***  &                      \\
WFH satisfaction: collaboration          & -0.477 (0.129)*** &                      &  & Change in car ownership              & 0.002 (0.074)     &                      \\
WFH satisfaction: feeling connected      & 0.179 (0.120)     &                      &  & Car ownership (fall 2021)            & 0.007 (0.044)     &                      \\
WFH satisfaction: work-life balance      & -0.002 (0.081)    &                      &  & Overall WFH satisfaction             & 0.025 (0.048)     &                      \\
Change in car ownership                  & 0.503 (0.188)***  &                      &  &                                      &                   &                      \\
Car ownership (fall 2021)                & 0.115 (0.114)     &                      &  &                                      &                   &                      \\ \bottomrule
\multicolumn{5}{l}{\textit{Note:} Standard errors are reported in parentheses. Dependent variables are denoted in bold. }  & \multicolumn{2}{r}{$^{*}$p$<$0.1; $^{**}$p$<$0.05; $^{***}$p$<$0.01} \\ 
\multicolumn{5}{l}{Log-transformation has been adopted to the usage frequency variables for all the modes.}  & \multicolumn{2}{r}{} \\ 
\multicolumn{5}{l}{Only the significant coefficients are reported.}  & \multicolumn{2}{r}{} \\ 
\end{tabular}%
} 
\end{table}
\FloatBarrier

\subsection{5.2 WFH prediction}
\begin{figure}[H]
    \centering
    \begin{minipage}{\textwidth} 
\includegraphics[width=0.9\textwidth]{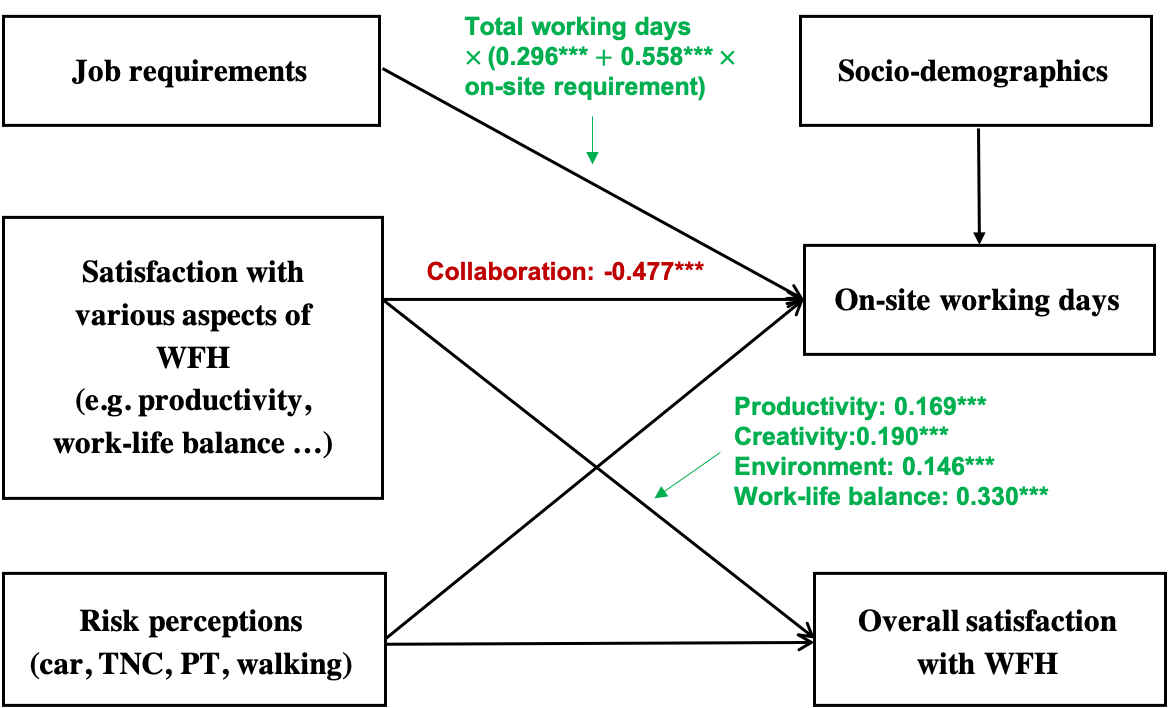}\centering
{\footnotesize \par Note: green indicates positive coefficients, whereas red indicates negative coefficients. No significant correlations between risk perceptions towards different travel modes and on-site working days/overall satisfaction with WFH have been found. \par}
\end{minipage}
    \caption{WFH prediction results}
    \label{fig:sem_output_wfh}
\end{figure}

Table \ref{result_wfh} reports the results for the predictions of on-site working days and overall satisfaction towards WFH. Figure \ref{fig:sem_output_wfh} presents coefficients of the key variables. 
\subsubsection{5.2.1 On-site working days}
\noindent When predicting the number of on-site working days, first we find that the number of total working days and the interaction between the number of total working days and on-site work requirements are crucial determinants. For those whose jobs require them to travel to specific job locations, one additional total working day is associated with 0.296+0.558=0.854 additional on-site working days. However, for those whose jobs do not require travel to specific job locations, one additional total working day is only associated with 0.296 additional on-site working days. ``Reimbursement`` is also an important factor influencing the number of on-site working days. Specifically, if the cost of remote work is reimbursed by the employer, on-site working days are reduced by 0.9 days on average. \\

\noindent Two additional findings are worth highlighting. First, the number of on-site working days is significantly affected by the change in car ownership (coeff=0.503, p$<$0.01), but not by car ownership itself. This result suggests a correlation between people buying cars during the pandemic and the increase of on-site working days, which probably indicates that the intention of buying cars during the pandemic for many people is to be capable of commuting to work by car. Second, it is found that among all aspects of WFH, only the evaluation of WFH regarding ``collaboration with colleagues'' significantly affects the number of on-site working days (coeff=-0.477, p$<$0.01). Specifically, when the satisfaction level of WFH regarding ``collaboration with colleagues'' increases by one degree, the average number of on-site working days decreases by 0.48 days. This significant coefficient indicates that lower satisfaction in terms of ``collaboration with colleagues'' when working from home would encourage people to work on-site more. Interestingly, though feeling towards ``collaboration with colleagues'' during WFH is the only WFH-related perspective that significantly affects the number of on-site working days, this factor does not significantly influence respondents' overall WFH satisfaction, which reflects an inconsistency between behavior and attitude regarding WFH. Given the emphasis of work collaborations in people's on-site working decisions, employers can encourage employees to work in the co-working spaces, as co-working spaces can allow employees to collaborate with colleagues while working closer to home \cite{williamson2021future}.


\subsubsection{5.2.2 Overall satisfaction with WFH}

\noindent We also investigate how overall WFH satisfaction is affected by attitudes towards different aspects of WFH. The results in Table \ref{result_wfh} shows that feelings towards ``work-life balance'' is the strongest predictor of overall satisfaction (coeff=0.33, p$<$0.01). Feelings towards ``creativity at work'', ``work environment'' and ``work productivity'' are also positively correlated with overall satisfaction (coeff=0.192, 0.144 and 0.171, respectively, p$<$0.01). Specifically, when the satisfaction levels of WFH regarding ``work-life balance'', ``creativity at work'', ``work environment'', and ``work productivity'' increase by 1 degree, the overall satisfaction level of WFH increases by 0.33, 0.19, 0.15, and 0.17 degrees, respectively. Perceptions towards ``focus on work'', ``collaboration with colleagues'' and ``feeling connected to colleagues'' do not have significant associations with overall satisfaction towards WFH.\\

\noindent Regarding the socio-demographic variables, we find that individuals living in larger households are less satisfied with WFH (coeff=-0.079, p$<$0.01). One potential reason is that in a large household, people have to cooperate more within their households as the usual divisions between work, home, and school become blurred \cite{biroli2021family}, and thus may create tension within households. Also, primary earners in large households may need to take on more family chores such as childcare during the work day, leading to lower satisfaction with WFH. Larger households are also likely to be noisier and create more distractions. 

\subsection{5.3 Commuting mode choice prediction}
\begin{figure}[!htbp]
    \centering
    \includegraphics[width=5 in]{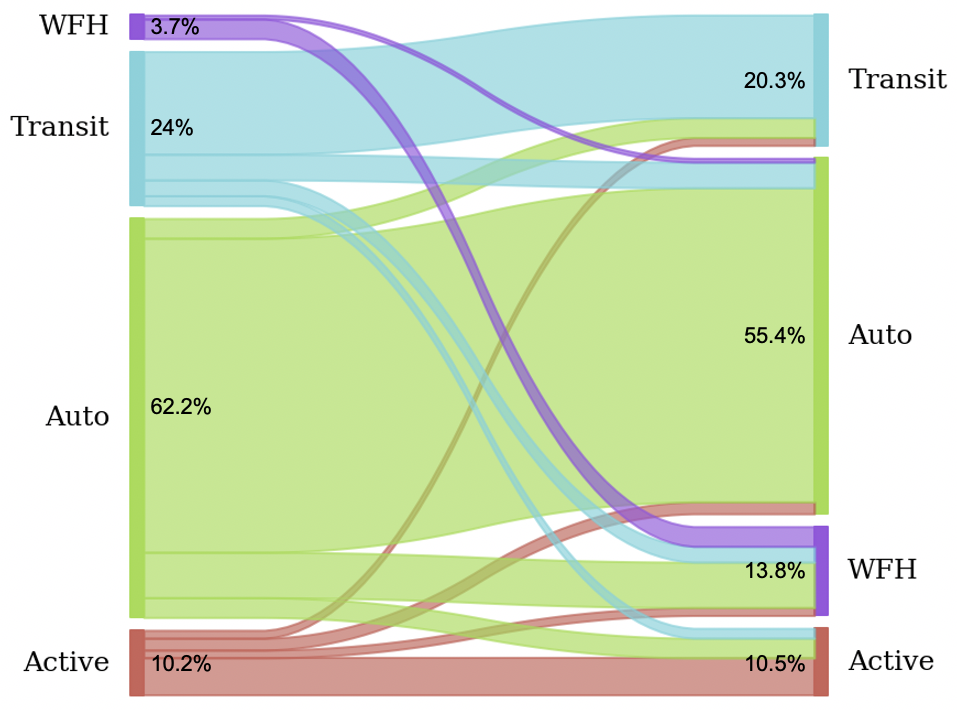}
    \caption{Mode shifts from the pre-pandemic time to the fall of 2021}
    \label{fig:mode_shift}
\end{figure}

\noindent We begin our analysis of commuting mode choice by presenting a Sankey Diagram that illustrates the shifts in transportation modes from the pre-pandemic period to the fall of 2021. As shown in Figure \ref{fig:mode_shift}, there have been significant changes in people's primary commuting modes during this period. To explore the factors that influence commuting mode choice in these two periods, we use discrete choice modeling and report our findings in Table \ref{result_DCM}. The likelihood ratio tests in the table compare the models for the pre-COVID period and the fall of 2021 against a base model reflecting market share (\textit{i.e.}, a model that only includes alternative-specific constants). The likelihood ratio test results indicate that the variables included in the model are jointly significant in explaining commuting mode choice for both the pre-COVID and the fall of 2021.\\

\noindent The results show that the travel time coefficients are significant for all travel modes in both the pre-pandemic time and the fall of 2021. Car ownership significantly impacts the utilities of commuting by car, transit, and work from home in the fall of 2021. In terms of the socio-demographic variables, higher income populations exhibit a positive relationship with commuting with active travel modes, while higher education is associated with smaller probability of walking compared with other commuting modes.\\

\noindent The analysis of parameter estimation does not disclose the extent to which the determinants influence individuals' travel mode choice. In order to quantify the magnitude of impact, it is crucial to estimate the elasticity effects of the explanatory variables. The elasticities of choosing different travel mode in response to the variations in travel time, car ownership, and income are reported in Table \ref{tab:elasticity}. Regarding travel time, it is observed that while the travel time elasticity for commuting by car does not change in these two periods, the absolute value of the travel time elasticity for active mode decreases substantially, from -11.052 to -7.992. In other words, a 1\% increase in travel time will increase the probability of walking for commute by 11\% in the pre-pandemic time, this figure shrinks to 8\% in the fall of 2021. The result indicates that people have become less sensitive to travel time when choosing to commute by walking. In contrast, the absolute elasticity of travel time for transit has increased from -0.487 to -0.534, showing that people have become more sensitive to the travel time of commuting by transit. The observed variations could be attributed to people's anxiety about contracting COVID-19 while commuting by transit for a relatively long travel time, coupled with a growing recognition of the benefits of walking in the fall of 2021.\\

\noindent Regarding the elasticity of various mode choices to car ownership, Table \ref{tab:elasticity} shows that as anticipated, an increase in car ownership is associated with a higher likelihood of commuting by car, and a lower probability of using active modes, public transit, and WFH. In terms of the magnitudes of the effects, we find that car ownership has a considerable impact on replacing active modes of transportation (elasticity = -2.643), but has a relatively minor impact on WFH (elasticity = -0.139).\\

\noindent Table \ref{tab:elasticity} also reveals interesting insights into the income effect on commuting behavior. We observe that high-income individuals are more likely to commute by active modes, and the elasticity of commuting by active modes with respect to income has significantly increased in the fall of 2021. In contrast, lower-income individuals tend to rely on cars for commuting, and the likelihood of car commutes among low-income individuals increases during the same period. This is probability because high-income individuals are more likely to embrace the benefits of active modes, such as environmental sustainability, improved health, and reduced risk of contracting COVID-19.


\FloatBarrier
\begin{table}[h]
\centering
  \caption{Commuting mode prediction results}
  \label{result_DCM}
\resizebox{0.9\textwidth}{!}{%
\begin{tabular}{@{}lll@{}}
\toprule
\textbf{Variable}                                & \textbf{Pre-COVID} & \textbf{Fall 2021} \\ \midrule
\multicolumn{3}{l}{\cellcolor[HTML]{D0CECE}Dependent variable}                             \\
\multicolumn{3}{l}{Primary commuting   mode to work/school: auto, transit, active, WFH}    \\
\multicolumn{3}{l}{\cellcolor[HTML]{D0CECE}Constants}                                      \\
ASC (auto)                                       & 2.025 (2.289)      & 6.712 (2.319)***   \\
ASC (transit)                                    & 1.614 (2.528)      & 5.101 (2.292)**    \\
ASC (WFH)                                        &                    & 5.350 (2.874)*     \\
\multicolumn{3}{l}{\cellcolor[HTML]{D0CECE}Alternative-specific   variables}               \\
Travel time (active)                             & -0.083 (0.025)***  & -0.055 (0.016)***  \\
Travel time (auto)                               & -0.093 (0.024)***  & -0.042 (0.017)**   \\
Travel time (transit)                            & -0.038 (0.009)***  & -0.025 (0.006)***  \\
WFH cost reimbursement (WFH)                     &                    & -0.297 (0.556)     \\
Satisfaction with WFH (WFH)                      &                    & 0.319 (0.253)      \\
\multicolumn{3}{l}{\cellcolor[HTML]{D0CECE}Socio-demographic   variables}                  \\
Car ownership (auto)                             & 2.863 (0.715)***   & 3.827 (0.782)***   \\
Car ownership (transit)                          & -0.015 (0.724)     & 1.715 (0.782)**    \\
Car ownership (WFH)                              &                    & 2.648 (1.079)**    \\
Age (auto)                                       & -0.041 (0.023)*    & -0.032 (0.029)     \\
Age (transit)                                    & -0.030 (0.028)     & -0.016 (0.032)     \\
Age (WFH)                                        &                    & -0.048 (0.039)     \\
Female (auto)                                    & 0.154 (0.594)      & 0.620 (0.734)      \\
Female (transit)                                 & 0.803 (0.660)      & 1.570 (0.771)**    \\
Female (WFH)                                     &                    & 1.513 (0.858)*     \\
Ethnicity: white (auto)                          & -1.639 (2.037)     & -2.268 (1.800)     \\
Ethnicity: white (transit)                       & -1.623 (2.142)     & -1.255 (1.846)     \\
Ethnicity: white (WFH)                           &                    & -3.512 (2.100)*    \\
Ethnicity: black or African American   (auto)    & -0.466 (2.198)     & -1.526 (1.892)     \\
Ethnicity: black or African American   (transit) & -1.412 (2.320)     & -0.685 (1.945)     \\
Ethnicity: black or African American   (WFH)     &                    & -1.820 (2.173)     \\
Ethnicity: Asian (auto)                          & -2.583 (2.102)     & -3.279 (2.001)     \\
Ethnicity: Asian (transit)                       & -2.780 (2.229)     & -2.367 (2.059)     \\
Ethnicity: Asian (WFH)                           &                    & -3.769 (2.293)     \\
Income (auto)                                    & -0.047 (0.039)     & -0.164 (0.066)**   \\
Income (transit)                                 & -0.034 (0.045)     & -0.139 (0.063)**   \\
Income (WFH)                                     &                    & -0.135 (0.070)*    \\
Education (auto)                                 & -0.061 (0.213)     & 1.007 (0.359)***   \\
Education (transit)                              & 0.225 (0.245)      & 1.235 (0.369)***   \\
Education (WFH)                                  &                    & 1.147 (0.422)***   \\
Household size (auto)                            & -0.021 (0.181)     & -0.221 (0.222)     \\
Household size (transit)                         & 0.218 (0.220)      & -0.011 (0.229)     \\
Household size (WFH)                             &                    & -0.292 (0.285)     \\
\multicolumn{3}{l}{\cellcolor[HTML]{D0CECE}Built environment   variables}                  \\
Distance to CBD (auto)                           & -0.032 (0.027)     & 0.011 (0.029)      \\
Distance to CBD (transit)                        & -0.043 (0.029)     & -0.010 (0.029)     \\
Distance to CBD (WFH)                            &                    & 0.000 (0.032)      \\
\multicolumn{3}{l}{\cellcolor[HTML]{D0CECE}Model fit}                                      \\
Sample size                                      & 349                & 335                \\
Final log likelihood                             & -176.55            & -222.31            \\
Likelihood ratio test      & 335.07***          & 339.81***         \\
McFadden pseudo-R$^{2}$       & 0.39               & 0.3                \\
Akaike Information Criterion                     & 419.1              & 546.61             \\
Bayesian Information Criterion                   & 546.32             & 741.13             \\ \bottomrule
\multicolumn{2}{l}{\textit{Note:} Robust standard errors in parentheses. For the pre-COVID period, }  & \multicolumn{1}{r}{$^{*}$p$<$0.1; $^{**}$p$<$0.05; $^{***}$p$<$0.01} \\ 
\multicolumn{2}{l}{only three alternatives (i.e. auto, transit, active) are included since there}  &  \\ 
\multicolumn{2}{l}{are too few WFH samples. }  &  \\ 
\end{tabular}%
}
\end{table}
\FloatBarrier

\begin{table}[h]
\centering
\caption{Elasticity effects}
\label{tab:elasticity}
\begin{tabular}{@{}lll@{}}
\toprule
\textbf{Variable}       & \textbf{Pre-COVID} & \textbf{Fall 2021} \\ \midrule
Travel time (active)    & -11.052            & -7.992             \\
Travel time (auto)      & -0.991             & -0.991             \\
Travel time (transit)   & -0.487             & -0.534             \\
Car ownership (active)  & -1.810             & -2.643             \\
Car ownership (transit) & -1.247             & -0.835             \\
Car ownership (auto)    & 0.683              & 0.700              \\
Car ownership (WFH)     & --               & -0.139             \\
Income (active)         & 0.349              & 1.189              \\
Income (transit)        & 0.039              & 0.032              \\
Income (auto)           & -0.077             & -0.239             \\
Income (WFH)            & --                & -0.013             \\ \bottomrule
\end{tabular}
\end{table}

\section{6. Discussion and conclusions}
\label{sec:conclusion}
The gradual return of pre-pandemic activities in the fall of 2021 as a result of the lifting of restrictions on dining, masks and indoor gatherings, can help us predict how the post-pandemic normal will look like. This research is one of the first attempts to explore the impact of COVID-19 on working from home and travel behavior in a U.S. city when COVID-19 restrictions were largely lifted and people were gradually returning to the office. We develop a comprehensive model of three travel choices (on-site working frequency, mode choice and car ownership) and study how these factors influence each other. We also examine what factors contribute to the number of on-site working days as well as satisfactions with WFH. The effects of attitudinal factors on travel choices and WFH decisions in the COVID-19 era are also examined. By accounting for pre-pandemic travel behavior, we identify the factors that contribute to shifts in travel behavior during the fall of 2021. Several key findings are summarized in the subsections below.\\

\subsection{6.1 Impact of COVID on travel mode usage frequencies}

Our online survey of Massachusetts residents reveals that the usage frequencies of PT and TNC have significantly reduced in the fall of 2021 when compared to pre-pandemic levels. Based on our SEM modeling results, we find that these reductions in PT and TNC ridership are partially affected by changes in car ownership, as an increase in car ownership leads to a 28\% decline in PT usage frequency, and an 15.9\% decline in TNC usage frequency. A smaller number of on-site working days are also associated with a lower usage frequency of PT, although this association does not exist between the number of on-site working days and other travel modes. This result reflects the negative impact of working from home on PT ridership. \\

\noindent The declines in usage frequencies of PT and TNC are also partially attributed to perceived risks of contracting COVID-19. Specifically, it is found that a lower perceived risk when using car decreases PT and TNC use. These results confirm that attitudinal factors play important roles in people's travel choices. \\

\subsection{6.2 Determinants of on-site working days}
Our SEM model also explores the influence of attitudinal factors on the number of on-site working days. The results show that the requirement to work on-site, the total working days and whether the employer reimburses the cost of remote work are crucial determinants of the number of on-site working days. These findings echo previous research that the ability to do a job remotely, employer support and availability and the cost of technology were the most important factors governing the intention to increase WFH \cite{jain2022covid,dey2020ability,kramer2020potential}. Therefore, if the local authorities want to sustain a behavior shift to WFH so as to alleviate traffic congestion, they should encourage employer support for WFH such as offering reimbursement for home office equipment. \\

\noindent Satisfaction towards WFH has also been found to contribute to WFH behavior. It is worth highlighting that the respondents' satisfaction with WFH regarding collaboration with their colleagues is the only WFH-related attitudinal factor that significantly affects the number of on-site working days. Therefore, if the local authorities or employers want to encourage remote working, they should develop measures to improve collaborations among employees, such as incentivizing employees to work in the co-working spaces located near employees' homes, which can enable employees to work together without having to commute long distance to work. \\

\noindent We also examine the relative importance of various aspects of WFH with respect to overall satisfaction towards WFH. We find that ``work-life balance'' is the most important aspect that influences general satisfaction, followed by ``creativity at work'', ``quality of the work environment'', and ``work productivity''. When working from home, respondents are less satisfied in terms of feeling connected to their colleagues and collaboration with their colleagues, but these elements are not found to have significant effects on determining their general satisfaction at work. Interestingly, given the importance of ``collaboration with colleagues'' in predicting the number of on-site working days, this factor is not significant when predicting respondents' overall WFH satisfaction. This result may reflect a disconnect between attitude and behavior regarding WFH.\\

\subsection{6.3 Commuting mode shifts}
This research also contributes to the existing literature by studying the potential shifts in commuting mode choice due to COVID-19 and the underlying factors that lead to the shifts. Since WFH is likely to be sustained to some extent in the post-pandemic era, it is critical to incorporate WFH into commuting mode choice models along with traditional travel mode alternatives  for the fall of 2021\cite{hensher2022impact}. Our survey result shows that the proportion of respondents that work from home full-time has increased from 3.6\% in the pre-pandemic time to 14.1\% in the fall of 2021. The discrete choice modeling results indicate that compared with the pre-pandemic period, people in the fall of 2021 are less sensitive to the travel time for active modes when choosing their commuting modes, but are more sensitive to the travel time for transit. Car ownership has a negative impact on the utility of WFH and the utilities of commuting by active modes and PT, and the effect is the most elastic regarding the active mode while is the least elastic regarding WFH. The elasticity of the probability of commuting by car to car ownership has increased in the fall of 2021. This finding coincides with the SEM modeling result showing that the change in car ownership positively correlates with the number of on-site working days. These results tell us that people have become more likely to use their cars to commute in the fall of 2021, and for those who have bought cars during the pandemic, they tend to work on-site more. In terms of socio-demographic heterogeneity in commuting mode choice, it is found that higher-income individuals are more likely to commute to work by active modes rather than commute by car, and this tendency has become more significant in the fall of 2021. This is probably due to the fact that higher-income individuals tend to appreciate the advantages of commuting by active modes such as environmental sustainability, improved health, and reduced risk of contracting COVID-19.\\

\subsection{6.4 Implications and future research}
This study has several limitations that call for future research. First, the fuel cost has been rising since the mid-2020 \cite{fuel2022}, which may impact car usage and ownership. However, since our study is based on a snapshot of travel behavior during a specific time period (i.e. the fall of 2021), we are not able to capture the impact of changes in fuel cost over time. We acknowledge that rising fuel costs may affect travel behavior and car ownership decisions in the future, and further research using longitudinal data would be necessary to investigate this effect. Second, the COVID-19 pandemic has led to a shortage in car supply, which could have affected both the ownership and usage of cars. This scarcity has affected not only the market for brand new cars, but also the second-hand market \cite{carsale2022,used_car2022}, which may have had an impact on individuals across different income groups. Further research could examine how the shortage in car supply has influenced car usage and ownership for people of varying income levels. \\

\noindent Overall, this research gathers valuable information on actual behavioral shifts in the fall of 2021 among residents of Greater Boston and provides insights on factors that drive these shifts. There is still much room for broader exploration of the research topics discussed in this paper. First, soon after our survey was completed, the COVID-19 Omicron variant spread around the country, which disrupted work and life once again. Now as the Omicron wave fades, it is imperative to explore whether the findings derived from this research have fluctuated. Second, while this research reveals changes in travel patterns and WFH behavior due to COVID and investigates what cause these changes, the authors do not take a stance on how decision makers should deal with these changes. In fact, how future policies might be designed to confront travel behavior shifts need further discussion, and more research needs to be done to analyze how different policies impact travel choices and WFH behavior in the post-pandemic era.

\newpage
\bibliographystyle{trb}
\bibliography{main}

\begin{thebibliography}{63}
\providecommand{\natexlab}[1]{#1}

\bibitem[{Mouratidis and Papagiannakis(2021)}]{mouratidis2021covid}
Mouratidis, K. and A.~Papagiannakis, COVID-19, internet, and mobility: The rise
  of telework, telehealth, e-learning, and e-shopping. \emph{Sustainable Cities
  and Society}, Vol.~74, 2021, p. 103182.

\bibitem[{Shakibaei et~al.(2021)Shakibaei, De~Jong, Alpk{\"o}kin, and
  Rashidi}]{shakibaei2021impact}
Shakibaei, S., G.~C. De~Jong, P.~Alpk{\"o}kin, and T.~H. Rashidi, Impact of the
  COVID-19 pandemic on travel behavior in Istanbul: A panel data analysis.
  \emph{Sustainable cities and society}, Vol.~65, 2021, p. 102619.

\bibitem[{Benita(2021)}]{benita2021human}
Benita, F., Human mobility behavior in COVID-19: A systematic literature review
  and bibliometric analysis. \emph{Sustainable Cities and Society}, Vol.~70,
  2021, p. 102916.

\bibitem[{Abdullah et~al.(2020)Abdullah, Dias, Muley, and
  Shahin}]{abdullah2020exploring}
Abdullah, M., C.~Dias, D.~Muley, and M.~Shahin, Exploring the impacts of
  COVID-19 on travel behavior and mode preferences. \emph{Transportation
  Research Interdisciplinary Perspectives}, Vol.~8, 2020, p. 100255.

\bibitem[{Mo et~al.(2021)Mo, Feng, Shen, Tam, Li, Yin, and
  Zhao}]{mo2021modeling}
Mo, B., K.~Feng, Y.~Shen, C.~Tam, D.~Li, Y.~Yin, and J.~Zhao, Modeling epidemic
  spreading through public transit using time-varying encounter network.
  \emph{Transportation Research Part C: Emerging Technologies}, Vol. 122, 2021,
  p. 102893.

\bibitem[{Jain et~al.(2022)Jain, Currie, and Aston}]{jain2022covid}
Jain, T., G.~Currie, and L.~Aston, COVID and working from home: Long-term
  impacts and psycho-social determinants. \emph{Transportation Research Part A:
  Policy and Practice}, Vol. 156, 2022, pp. 52--68.

\bibitem[{Levin et~al.(1977)Levin, Mosell, Lamka, Savage, and
  Gray}]{levin1977measurement}
Levin, I.~P., M.~K. Mosell, C.~M. Lamka, B.~E. Savage, and M.~Gray, Measurement
  of psychological factors and their role in travel behavior. \emph{Choice},
  Vol.~4, No.~11, 1977, pp. 15--6.

\bibitem[{Zheng et~al.(2022)Zheng, Kong, Petzhold, Barcelos, Zegras, and
  Zhao}]{zheng2022gender}
Zheng, Y., H.~Kong, G.~Petzhold, M.~M. Barcelos, C.~P. Zegras, and J.~Zhao,
  Gender differences in the user satisfaction and service quality improvement
  priority of public transit bus system in Porto Alegre and Fortaleza, Brazil.
  \emph{Travel Behaviour and Society}, Vol.~28, 2022, pp. 22--37.

\bibitem[{Zheng et~al.(2021{\natexlab{a}})Zheng, Kong, Petzhold, Barcelos,
  Zegras, and Zhao}]{zheng2021user}
Zheng, Y., H.~Kong, G.~Petzhold, M.~M. Barcelos, C.~P. Zegras, and J.~Zhao,
  User satisfaction and service quality improvement priority of bus rapid
  transit in Belo Horizonte, Brazil. \emph{Case Studies on Transport Policy},
  Vol.~9, No.~4, 2021{\natexlab{a}}, pp. 1900--1911.

\bibitem[{Golob(2003)}]{golob2003structural}
Golob, T.~F., Structural equation modeling for travel behavior research.
  \emph{Transportation Research Part B: Methodological}, Vol.~37, No.~1, 2003,
  pp. 1--25.

\bibitem[{Van~Acker et~al.(2007)Van~Acker, Witlox, and
  Van~Wee}]{van2007effects}
Van~Acker, V., F.~Witlox, and B.~Van~Wee, The effects of the land use system on
  travel behavior: A structural equation modeling approach.
  \emph{Transportation Planning and Technology}, Vol.~30, No.~4, 2007, pp.
  331--353.

\bibitem[{Bagley and Mokhtarian(2002)}]{bagley2002impact}
Bagley, M.~N. and P.~L. Mokhtarian, The impact of residential neighborhood type
  on travel behavior: A structural equations modeling approach. \emph{The
  Annals of Regional Science}, Vol.~36, No.~2, 2002, pp. 279--297.

\bibitem[{De~Vos et~al.(2021)De~Vos, Cheng, and Witlox}]{de2021changes}
De~Vos, J., L.~Cheng, and F.~Witlox, Do changes in the residential location
  lead to changes in travel attitudes? A structural equation modeling approach.
  \emph{Transportation}, Vol.~48, No.~4, 2021, pp. 2011--2034.

\bibitem[{Etminani-Ghasrodashti and Ardeshiri(2015)}]{etminani2015modeling}
Etminani-Ghasrodashti, R. and M.~Ardeshiri, Modeling travel behavior by the
  structural relationships between lifestyle, built environment and non-working
  trips. \emph{Transportation Research Part A: Policy and Practice}, Vol.~78,
  2015, pp. 506--518.

\bibitem[{Moody and Zhao(2020)}]{moody2020travel}
Moody, J. and J.~Zhao, Travel behavior as a driver of attitude: Car use and car
  pride in US cities. \emph{Transportation Research Part F: Traffic Psychology
  and Behaviour}, Vol.~74, 2020, pp. 225--236.

\bibitem[{Gao et~al.(2020)Gao, Rao, Kang, Liang, and Kruse}]{gao2020mapping}
Gao, S., J.~Rao, Y.~Kang, Y.~Liang, and J.~Kruse, Mapping county-level mobility
  pattern changes in the United States in response to COVID-19.
  \emph{SIGSpatial Special}, Vol.~12, No.~1, 2020, pp. 16--26.

\bibitem[{Heiler et~al.(2020)Heiler, Reisch, Hurt, Forghani, Omani, Hanbury,
  and Karimipour}]{heiler2020country}
Heiler, G., T.~Reisch, J.~Hurt, M.~Forghani, A.~Omani, A.~Hanbury, and
  F.~Karimipour, Country-wide mobility changes observed using mobile phone data
  during COVID-19 pandemic. In \emph{2020 IEEE International Conference on Big
  Data (Big Data)}, IEEE, 2020, pp. 3123--3132.

\bibitem[{Xiong et~al.(2020)Xiong, Hu, Yang, Younes, Luo, Ghader, and
  Zhang}]{xiong2020mobile}
Xiong, C., S.~Hu, M.~Yang, H.~Younes, W.~Luo, S.~Ghader, and L.~Zhang, Mobile
  device location data reveal human mobility response to state-level
  stay-at-home orders during the COVID-19 pandemic in the USA. \emph{Journal of
  the Royal Society Interface}, Vol.~17, No. 173, 2020, p. 20200344.

\bibitem[{Shamshiripour et~al.(2020)Shamshiripour, Rahimi, Shabanpour, and
  Mohammadian}]{shamshiripour2020covid}
Shamshiripour, A., E.~Rahimi, R.~Shabanpour, and A.~K. Mohammadian, How is
  COVID-19 reshaping activity-travel behavior? Evidence from a comprehensive
  survey in Chicago. \emph{Transportation Research Interdisciplinary
  Perspectives}, Vol.~7, 2020, p. 100216.

\bibitem[{Barrero et~al.(2021)Barrero, Bloom, and Davis}]{barrero2021working}
Barrero, J.~M., N.~Bloom, and S.~J. Davis, \emph{Why working from home will
  stick}. National Bureau of Economic Research, 2021.

\bibitem[{Currie et~al.(2021)Currie, Jain, and Aston}]{currie2021evidence}
Currie, G., T.~Jain, and L.~Aston, Evidence of a post-COVID change in travel
  behaviour: Self-reported expectations of commuting in Melbourne.
  \emph{Transportation Research Part A: Policy and Practice}, Vol. 153, 2021,
  pp. 218--234.

\bibitem[{Abdullah et~al.(2021)Abdullah, Ali, Hussain, Aslam, and
  Javid}]{abdullah2021measuring}
Abdullah, M., N.~Ali, S.~A. Hussain, A.~B. Aslam, and M.~A. Javid, Measuring
  changes in travel behavior pattern due to COVID-19 in a developing country: A
  case study of Pakistan. \emph{Transport Policy}, Vol. 108, 2021, pp. 21--33.

\bibitem[{Torrisi et~al.(2021)Torrisi, Campisi, Inturri, Ignaccolo, and
  Tesoriere}]{torrisi2021continue}
Torrisi, V., T.~Campisi, G.~Inturri, M.~Ignaccolo, and G.~Tesoriere, Continue
  to share? An overview on italian travel behavior before and after the
  COVID-19 lockdown. In \emph{AIP Conference Proceedings}, AIP Publishing LLC,
  2021, Vol. 2343, p. 090010.

\bibitem[{Anwari et~al.(2021)Anwari, Ahmed, Islam, Hadiuzzaman, and
  Amin}]{anwari2021exploring}
Anwari, N., M.~T. Ahmed, M.~R. Islam, M.~Hadiuzzaman, and S.~Amin, Exploring
  the travel behavior changes caused by the COVID-19 crisis: A case study for a
  developing country. \emph{Transportation Research Interdisciplinary
  Perspectives}, Vol.~9, 2021, p. 100334.

\bibitem[{Parady et~al.(2020)Parady, Taniguchi, and Takami}]{parady2020travel}
Parady, G., A.~Taniguchi, and K.~Takami, Travel behavior changes during the
  COVID-19 pandemic in Japan: Analyzing the effects of risk perception and
  social influence on going-out self-restriction. \emph{Transportation Research
  Interdisciplinary Perspectives}, Vol.~7, 2020, p. 100181.

\bibitem[{Beck et~al.(2020)Beck, Hensher, and Wei}]{beck2020slowly}
Beck, M.~J., D.~A. Hensher, and E.~Wei, Slowly coming out of COVID-19
  restrictions in Australia: Implications for working from home and commuting
  trips by car and public transport. \emph{Journal of Transport Geography},
  Vol.~88, 2020, p. 102846.

\bibitem[{Barbour et~al.(2021)Barbour, Menon, and
  Mannering}]{barbour2021statistical}
Barbour, N., N.~Menon, and F.~Mannering, A statistical assessment of
  work-from-home participation during different stages of the COVID-19
  pandemic. \emph{Transportation Research Interdisciplinary Perspectives},
  Vol.~11, 2021, p. 100441.

\bibitem[{Aaditya and Rahul(2021)}]{aaditya2021psychological}
Aaditya, B. and T.~Rahul, Psychological impacts of COVID-19 pandemic on the
  mode choice behaviour: A hybrid choice modelling approach. \emph{Transport
  Policy}, Vol. 108, 2021, pp. 47--58.

\bibitem[{Chen et~al.(2021)Chen, Feng, and Gu}]{chen2021role}
Chen, C., T.~Feng, and X.~Gu, Role of latent factors and public policies in
  travel decisions under COVID-19 pandemic: Findings of a hybrid choice model.
  \emph{Sustainable Cities and Society}, 2021, p. 103601.

\bibitem[{Irawan et~al.(2021)Irawan, Belgiawan, Joewono, Bastarianto, Rizki,
  and Ilahi}]{irawan2021exploring}
Irawan, M.~Z., P.~F. Belgiawan, T.~B. Joewono, F.~F. Bastarianto, M.~Rizki, and
  A.~Ilahi, Exploring activity-travel behavior changes during the beginning of
  COVID-19 pandemic in Indonesia. \emph{Transportation}, 2021, pp. 1--25.

\bibitem[{Javid et~al.(2021)Javid, Abdullah, Ali, and
  Dias}]{javid2021structural}
Javid, M.~A., M.~Abdullah, N.~Ali, and C.~Dias, Structural equation modeling of
  public transport use with COVID-19 precautions: An extension of the norm
  activation model. \emph{Transportation Research Interdisciplinary
  Perspectives}, Vol.~12, 2021, p. 100474.

\bibitem[{Elias and Zatmeh-Kanj(2021)}]{elias2021extent}
Elias, W. and S.~Zatmeh-Kanj, Extent to which COVID-19 will affect future use
  of the train in Israel. \emph{Transport Policy}, Vol. 110, 2021, pp.
  215--224.

\bibitem[{Jiao et~al.(2021)Jiao, Bhat, and Azimian}]{jiao2021measuring}
Jiao, J., M.~Bhat, and A.~Azimian, Measuring travel behavior in Houston, Texas
  with mobility data during the 2020 COVID-19 outbreak. \emph{Transportation
  Letters}, Vol.~13, No. 5-6, 2021, pp. 461--472.

\bibitem[{Asmussen et~al.(2022)Asmussen, Mondal, Bhat, and
  Pendyala}]{asmussen2022modeling}
Asmussen, K.~E., A.~Mondal, C.~R. Bhat, and R.~M. Pendyala, {On modeling future
  workplace location decisions: An analysis of Texas employees}.
  \emph{Technical paper, Department of Civil, Architectural and Environmental
  Engineering, The University of Texas at Austin}, 2022.

\bibitem[{Salon et~al.(2021)Salon, Conway, Capasso~da Silva, Chauhan, Derrible,
  Mohammadian, Khoeini, Parker, Mirtich, Shamshiripour
  et~al.}]{salon2021potential}
Salon, D., M.~W. Conway, D.~Capasso~da Silva, R.~S. Chauhan, S.~Derrible,
  A.~Mohammadian, S.~Khoeini, N.~Parker, L.~Mirtich, A.~Shamshiripour, et~al.,
  {The potential stickiness of pandemic-induced behavior changes in the United
  States}. \emph{Proceedings of the National Academy of Sciences}, Vol. 118,
  No.~27, 2021, p. e2106499118.

\bibitem[{Tahlyan et~al.(2022)Tahlyan, Said, Mahmassani, Stathopoulos, Shaheen,
  Walker et~al.}]{tahlyan2022longitudinal}
Tahlyan, D., M.~Said, H.~Mahmassani, A.~Stathopoulos, S.~Shaheen, J.~L. Walker,
  et~al., \emph{Longitudinal tracking survey to understand changing consumer
  spending, telework and mobility patterns through the pandemic}. Tier-1
  University Transportation Center on Telemobility (Telemobility UTC), 2022.

\bibitem[{Kong et~al.(2022)Kong, Zhang, Xiao, Das, and Zhang}]{kong2022work}
Kong, X., A.~Zhang, X.~Xiao, S.~Das, and Y.~Zhang, Work from home in the
  post-COVID world. \emph{Case Studies on Transport Policy}, Vol.~10, No.~2,
  2022, pp. 1118--1131.

\bibitem[{{Massachusetts Department of Public Health}(2020)}]{massdph2020}
{Massachusetts Department of Public Health}, \emph{{Man returning from Wuhan,
  China is first case of 2019 Novel Coronavirus confirmed in Massachusetts}}.
  Online,
  \url{https://www.mass.gov/news/man-returning-from-wuhan-china-is-first-case-of-2019-novel-coronavirus-confirmed-in-massachusetts},
  2020.

\bibitem[{Arsenault(2020)}]{arsenault2020biogen}
Arsenault, M., {How the Biogen leadership conference in Boston spread the
  coronavirus}. \emph{Boston Globe}, 2020.

\bibitem[{Reiss et~al.(2020)Reiss, Greenberg, Sweeney, Kaufman, Enos, Feldman,
  Patel, and Purifoy}]{reiss2020schools}
Reiss, J., Z.~Greenberg, E.~Sweeney, A.~Kaufman, C.~Enos, A.~Feldman, S.~Patel,
  and S.~Purifoy, {A list of Mass. school closures amid coronavirus outbreak}.
  \emph{Boston Globe}, 2020.

\bibitem[{Markos(2021)}]{markos2021lockdown}
Markos, M., {Life in lockdown: A timeline of the COVID shutdown in
  Massachusetts}. \emph{NBC Boston}, 2021.

\bibitem[{Hilliard(2020)}]{hilliard2020reopening}
Hilliard, J., {Boston enters Phase 3 of reopening today. But business owners
  aren’t expecting a flood of customers}. \emph{Boston Globe}, 2020.

\bibitem[{Stout and Moore(2020)}]{stout2020mandate}
Stout, M. and D.~Moore, {Governor Baker announces tighter mask mandate, other
  COVID-19 restrictions}. \emph{Boston Globe}, 2020.

\bibitem[{Platoff and Prignano(2021)}]{platoff2021policies}
Platoff, E. and C.~Prignano, {Baker says he plans no changes in COVID-19
  policies amid a rise in cases}. \emph{Boston Globe}, 2021.

\bibitem[{Carlin(2021)}]{carlin2021college}
Carlin, J., {`Slowly going back to normality': College students return to
  campus in Boston}. \emph{Boston Globe}, 2021.

\bibitem[{Gardizy(2021)}]{gardizy2021office}
Gardizy, A., {There are reasons to still avoid the office, but health concerns
  aren’t necessarily among them}. \emph{Boston Globe}, 2021.

\bibitem[{Bleichfield(2021)}]{bleichfield2021mbta}
Bleichfield, A., {MBTA ridership rebounds, somewhat, from pandemic}. \emph{The
  Bay State Banner}, 2021.

\bibitem[{Hair(2009)}]{hair2009multivariate}
Hair, J.~F., \emph{Multivariate data analysis: A global perspective}.
  Prentice-Hall, 7th ed., 2009.

\bibitem[{Hair~Jr et~al.(2021)Hair~Jr, Hult, Ringle, and
  Sarstedt}]{hair2021primer}
Hair~Jr, J.~F., G.~T.~M. Hult, C.~M. Ringle, and M.~Sarstedt, \emph{A primer on
  partial least squares structural equation modeling (PLS-SEM)}. Sage
  Publications, 2021.

\bibitem[{Moody et~al.(2021)Moody, Farr, Papagelis, and Keith}]{moody2021value}
Moody, J., E.~Farr, M.~Papagelis, and D.~R. Keith, The value of car ownership
  and use in the United States. \emph{Nature Sustainability}, 2021, pp. 1--6.

\bibitem[{Hensher et~al.(2022)Hensher, Balbontin, Beck, and
  Wei}]{hensher2022impact}
Hensher, D.~A., C.~Balbontin, M.~J. Beck, and E.~Wei, The impact of working
  from home on modal commuting choice response during COVID-19: Implications
  for two metropolitan areas in Australia. \emph{Transportation Research Part
  A: Policy and Practice}, Vol. 155, 2022, pp. 179--201.

\bibitem[{Ajzen(1991)}]{ajzen1991theory}
Ajzen, I., The theory of planned behavior. \emph{Organizational Behavior and
  Human Decision Processes}, Vol.~50, No.~2, 1991, pp. 179--211.

\bibitem[{Ben-Akiva et~al.(1985)Ben-Akiva, Lerman, Lerman
  et~al.}]{ben1985discrete}
Ben-Akiva, M.~E., S.~R. Lerman, S.~R. Lerman, et~al., \emph{Discrete choice
  analysis: Theory and application to travel demand}, Vol.~9. MIT press, 1985.

\bibitem[{Zheng et~al.(2021{\natexlab{b}})Zheng, Wang, and
  Zhao}]{zheng2021equality}
Zheng, Y., S.~Wang, and J.~Zhao, Equality of opportunity in travel behavior
  prediction with deep neural networks and discrete choice models.
  \emph{Transportation Research Part C: Emerging Technologies}, Vol. 132,
  2021{\natexlab{b}}, p. 103410.

\bibitem[{Ullman and Bentler(2012)}]{ullman2012structural}
Ullman, J.~B. and P.~M. Bentler, Structural equation modeling. \emph{Handbook
  of Psychology, Second Edition}, Vol.~2, 2012.

\bibitem[{Zheng et~al.(2023)Zheng, Meredith-Karam, Stewart, Kong, and
  Zhao}]{zheng2023impacts}
Zheng, Y., P.~Meredith-Karam, A.~Stewart, H.~Kong, and J.~Zhao, Impacts of
  congestion pricing on ride-hailing ridership: Evidence from Chicago.
  \emph{Transportation Research Part A: Policy and Practice}, Vol. 170, 2023,
  p. 103639.

\bibitem[{Williamson et~al.(2021)Williamson, Pearce, Dickinson, Weeratunga, and
  Bucknall}]{williamson2021future}
Williamson, S., A.~Pearce, H.~Dickinson, V.~Weeratunga, and F.~Bucknall,
  \emph{Future of work literature review: Emerging trends and issues}. Public
  Service Research Group, 2021.

\bibitem[{Biroli et~al.(2021)Biroli, Bosworth, Della~Giusta, Di~Girolamo,
  Jaworska, and Vollen}]{biroli2021family}
Biroli, P., S.~Bosworth, M.~Della~Giusta, A.~Di~Girolamo, S.~Jaworska, and
  J.~Vollen, Family life in lockdown. \emph{Frontiers in Psychology}, Vol.~12,
  2021.

\bibitem[{Dey et~al.(2020)Dey, Frazis, Loewenstein, and Sun}]{dey2020ability}
Dey, M., H.~Frazis, M.~A. Loewenstein, and H.~Sun, Ability to work from home.
  \emph{Monthly Labor Review}, 2020, pp. 1--19.

\bibitem[{Kramer and Kramer(2020)}]{kramer2020potential}
Kramer, A. and K.~Z. Kramer, The potential impact of the COVID-19 pandemic on
  occupational status, work from home, and occupational mobility. \emph{Journal
  of Vocational Behavior}, Vol. 119, 2020, p. 103442.

\bibitem[{{Bureau of Transportation Statistics}(2022)}]{fuel2022}
{Bureau of Transportation Statistics}, \emph{{Record Breaking Increases in
  Motor Fuel Prices in 2022}}. Online,
  \url{https://www.bts.gov/data-spotlight/record-breaking-increases-motor-fuel-prices-2022},
  2022.

\bibitem[{{AP News}(2022)}]{carsale2022}
{AP News}, \emph{{New auto sales up in 2021, but long way before full
  recovery}}. Online,
  \url{https://apnews.com/article/coronavirus-pandemic-business-health-3e51e0b4dd6afac30a21aa6c3725c05b},
  2022.

\bibitem[{{New York Times}(2023)}]{used_car2022}
{New York Times}, \emph{{The Pandemic Used-Car Boom Is Coming to an Abrupt
  End}}. Online,
  \url{https://www.nytimes.com/2023/01/30/business/economy/used-cars-carmax-carvana.html},
  2023.

\end{thebibliography}
\nolinenumbers
\end{document}